# Resolving ultrafast exciton migration in organic solids at the nanoscale

Samuel B. Penwell[1†], Lucas D. S. Ginsberg[1], Rodrigo Noriega[1‡], and Naomi S. Ginsberg[1,2,3,4,5*]

[1]Department of Chemistry, University of California, Berkeley, California 94720, United States
[2]Molecular Biophysics and Integrative Bioimaging Division, Lawrence Berkeley National Laboratory, Berkeley, California, 94720, United States
[3]Materials Science Division, Lawrence Berkeley National Laboratory, Berkeley, California, 94720, United States
[4]Kavli Energy NanoScience Institute, Berkeley, California 94720, United States
[5]Department of Physics, University of California, Berkeley, California 94720, United States
[*]Correspondence to: nsginsberg@berkeley.edu
[†]Current address: James Franck Institute, University of Chicago, Chicago, IL 60615, United States
[‡]Current address: Department of Chemistry, University of Utah, Salt Lake City, Utah 84112, United States

**The effectiveness of molecular-based light harvesting relies on transport of optical excitations, excitons, to charge-transfer sites.**[1–7] **Measuring exciton migration**[3,7–16] **has, however, been challenging because of the mismatch between nanoscale migration lengths and the diffraction limit. In organic semiconductors, common bulk methods employ a series of films terminated at quenching substrates, altering the spatioenergetic landscape for migration.**[9,12,13,16] **Here we instead define quenching boundaries all-optically with sub-diffraction resolution, thus characterizing spatiotemporal exciton migration on its native nanometer and picosecond scales without disturbing morphology. By transforming stimulated emission depletion microscopy into a time-resolved ultrafast approach, we measure a 16-nm migration length in CN-PPV conjugated polymer films. Combining these experiments with Monte Carlo exciton hopping simulations shows that migration in CN-PPV films is essentially diffusive because intrinsic chromophore energetic disorder is comparable to inhomogeneous broadening among chromophores. This framework also illustrates general trends across materials. Our new approach's sub-diffraction resolution will enable previously unattainable correlations of local material structure to the nature of exciton migration, applicable not only to photovoltaic or display-destined organic semiconductors but also to explaining the quintessential exciton migration exhibited in photosynthesis.**

Complex materials with electronically coupled chromophores, such as those in organic or composite nanocrystal photovoltaics and light emitting diodes, or in photosynthetic light harvesting, rely on the migration of tightly-bound, localized, short-lived excitons for energy transduction.[1–3,5–7,17,18] Exciton migration in photosynthesis is nearly 100% efficient, but the design principles enabling efficient transport in such complex and disordered systems have remained challenging.[17,18] Conversely, organic photovoltaic performance has been limited by the short, ~5-20 nm-range of exciton migration in organic semiconductors.[3,7–13] There has been considerable effort to understand the origin of these limited migration lengths, but the development of a complete physical picture of the migration process has been hindered by the challenge associated with measuring exciton migration and correlating it to the local, disordered material substructure.[19–24] Since excitons are optical excitations, spatial resolution in migration measurements is generally limited by diffraction to >200 nm, even though migration lengths are 1-2 orders of magnitude smaller. Typically, this challenge is addressed by depositing a set of thin films with thicknesses on the order of the migration length onto exciton-quenching substrates. The length scale of exciton migration is then estimated from the dependence of the resulting photoluminescence quenching on film thickness.[3,9,12,13] These measurements, however, do not necessarily represent exciton migration in the absence of the quenching layer and are limited in their ability to discern heterogeneity since they average over areas larger than the scales of heterogeneity and require measuring multiple sample preparations.[3,9,16] Recently, an elegant new method was demonstrated that removes the need for a quenching boundary by resolving, over the nanoseconds time scale of fluorescence decay, the very small expansion of an excitation volume prepared at the diffraction limit.[14,15] Here we present an approach to surmount both the diffraction limit and the need for a physical quenching layer by employing an optical quenching boundary. Surmounting the diffraction limit leads to probe volumes with larger perimeter-to-area ratios so that migration near the perimeter forms a larger fraction of our signal and should also enable investigation of exciton migration heterogeneity on the scales of physical heterogeneity.

**Measuring exciton migration.** We have measured exciton migration in organic semiconducting conjugated polymer films on its native nanometer and picosecond scales. We did so by producing an ultrafast optical exciton quenching boundary, defined through spatially varying light-matter interactions on the characteristic scales of migration. More specifically, this feat first required that we extend photophysically restrictive stimulated emission depletion (STED) microscopy to electronically coupled materials with endogenous chromophores.[25] It furthermore required that we devise a means to use the sharp, sub-diffraction excited-state population profile thus created to both induce and measure the spatiotemporal evolution of the exciton distribution. STED microscopy is a form of super-resolution fluorescence imaging in which a pump laser pulse creates a diffraction-limited excitation distribution and a depletion laser pulse, with an annular transverse mode, drives stimulated emission around the periphery of the excitation distribution.[26] Only the small fraction of excitations at the center of the original distribution survives, defining a sub-diffraction excitation volume that can in principle be vanishingly small due to the intensity-dependent depletion saturation of the sample. As shown in our time-resolved ultrafast STED (TRUSTED) scheme in **Figure 1a frames 1&2,** we use this combination of a pump and an annular depletion pulse to define a sub-diffraction excitation volume of <100 nm (**Supplementary Figure 1**), but it is only able to serve as an initial condition for migration. Tracking the nanoscale exciton redistribution over its very brief lifetime requires additional innovation because emitted light collected in the far field does not encode information about the specific location (or spatial distribution) of emission and its evolution from the initial excitation profile.

We respond to this challenge by following the preparation of the excitation volume with a second annular depletion pulse. This pulse defines a subsequent sub-diffraction *detection* volume that produces a coaxial quenching boundary at a controllable and variable time delay (**Figure 1a frame 4**). The sharply confined sub-diffraction excitation volume initially has steep gradients at its boundary, inducing larger subsequent relative changes to the distribution than are achievable otherwise. By using the second depletion pulse to redefine a quenching boundary identical to the initial one, excitons that migrate beyond this boundary in the intervening period may be quenched. The extent of migration over time can thus be resolved by progressively increasing this intervening migration period over successive measurements and collecting fluorescence emission only from the correspondingly decreasing population remaining in the resulting 'detection volume' at the instant this second depletion pulse is applied (**Figure 1a frame 5**). The length and time scales of migration are encoded in the decay of the observed fluorescence signal as a function of this delay time. This TRUSTED approach enables us to measure migration on its native spatial and temporal scales in the absence of a physical quenching boundary, thereby opening the possibility for correlative measurements with the local material morphology in order to develop an understanding of the structure/function relationship that governs exciton migration.

We employ the TRUSTED scheme on poly(2,5-di(hexyloxy)cyanoterephthalylidene) (CN-PPV) thin films in a home-built confocal microscope (**Supplementary Figure 2 and text**), using a 3-fJ few-ps pump pulse at 540 nm and two identical 240-pJ ~120-ps depletion pulses at 740 nm (point spread functions shown in Supplementary **Figure 3**). The film has a 5-ns fluorescence lifetime[25,27], $\tau$, and representative absorption and emission spectra are shown in **Figure 1b**. We collect fluorescence between 635 and 640 nm and scan the delay time between excitation and detection from 240 ps to 3 ns. Care is taken to avoid sample heating by using transverse rastering. Isolating the component of the measured fluorescence decay that is due to migration requires great care. To do so, we first modulate the 540-nm excitation to eliminate the contribution of any two photon absorption (2PA) of the depletion pulses to the detected migration signal (see

**Supplementary text** for details).[25] Second, fluorescence detection using a single photon counting avalanche photodiode (SPAD) is gated on within <200 ps[28] approximately 500 ps after the arrival of the second depletion pulse to reject any fluorescence emitted prior to the definition of the detection volume.[29] Last, we normalize the fluorescence signal to that obtained in the absence of the second depletion pulse. This normalization removes the contribution to the decaying signal due to fluorescence decay of excitations and isolates the contribution due to the outflux of excitons from the initially excited volume. (See **Figure 1c** for the modulation scheme, **Supplementary Table 1**, **Supplementary Figure 4** and accompanying text for details.) The resulting normalized detection volume fluorescence can be directly related to the fraction of the exciton population at a given time that survives the definition of the detection volume. Its decay can be attributed to the increase in the fraction of the population that crosses and is quenched beyond the relatively abrupt "boundary" of this volume as a function of delay time, as simulated in **Figure 2b**. The observed fraction of remaining excitons versus delay time for a representative CN-PPV thin film is shown in **Figure 2a**. The curve overlaying the data is a fit that assumes diffusive exciton migration dynamics and includes the action of both spatially dependent depletion pulses (**Figure 2b**), yielding a migration length of $L_d$ = 16 ± 2 nm (**Supplementary Figures 5** and **6** and associated text). This migration length is defined for consistency with previous literature as $L_d = \sqrt{D\tau}$, for a diffusivity $D$ (see **Supplementary text**). This result is reproducible (**Supplementary Figure 7**) and the measured migration length is independent of the excitation density (**Supplementary Figure 8**), confirming that higher order effects such as exciton-exciton annihilation do not contribute.

**Modeling exciton migration.** Having succeeded in obtaining a local measure of the extent of exciton migration, we turn to explaining the relatively long $L_d$ for CN-PPV and its microscopic origin by employing Monte Carlo exciton hopping simulations[3,6,15,30–32] (see **Supplementary text**). We employ an incoherent hopping model to obtain qualitative trends in the below-defined parameter space because we expect incoherent transport over the time scales of import in our measurements. By more generally identifying and relating parameters that impact exciton migration, we also hypothesize how other electronically-coupled systems are situated within the same framework. Microscopically, we identify two distinct energetic considerations relevant to respectively determining the spatioenergetic landscape for migration and the way that excitons dynamically traverse it. First, a disordered organic solid is composed of closely spaced chromophore sites, where we define a site as the region over which an excitonic eigenstate is delocalized. The site transition energies are distributed due to variation in π-conjugation and configurations of moieties, interactions with neighboring chromophores, and variation in the local electrostatic environment. This inhomogeneous site energy distribution, with characteristic width $\sigma_{IH}$, defines the spatioenergetic landscape for exciton migration (**Supplementary Figure 9**). As important as defining this landscape for exciton migration is the energy scale that determines the way that the excitons are able to explore it. Therefore, we secondly consider the intrinsic excited-state configurational energy relaxation, or reorganization energy, of a chromophore, which determines the intramolecular Stokes shift, $\Delta$, that we find dominates the dynamic character of migration over the spatioenergetic landscape (**Supplementary Figure 9**). This intrinsic chromophore Stokes shift governs the typical exciton hop probability and duration by biasing excitations to hop downhill in energy and by limiting the spectral overlap for isoenergetic hops. In particular, whether exciton migration is diffusive or subdiffusive depends on the size of $\Delta$ relative to the total energy lost over the course of a migration trajectory, $\Delta E$. The effects of each of $\sigma_{IH}$ and $\Delta$ must, however, be considered relative to the intrinsic variation in a given chromophore's site energy, $\sigma_{IT}$. This intrinsic variation can be thought of as a generalized homogeneous broadening, occurring due to Franck-Condon progressions and fast thermal fluctuations, which for simplicity we take to be identical for each site. Since we find that $\sigma_{IT}$ establishes an effective energy resolution for these two other contributors to respectively determining the spatioenergetic landscape and the nature of migration over the landscape, we plot and classify the average output parameters of the simulated trajectories as a function of the interchromophore variable $\sigma_{IH}/\sigma_{IT}$, the effective distinguishability of sites from one another, and the intrachromophore variable $\Delta/\sigma_{IT}$, the strength of downhill bias for hopping within the inhomogeneous distribution.

By implementing this approach, we first determine the spatioenergetic landscape and corresponding nature of exciton migration that is most likely to underlie our $L_d$~16 nm TRUSTED result in **Figure 2**. We perform the Monte Carlo hopping simulations at multiple points in the $\sigma_{IH}/\sigma_{IT}$—$\Delta/\sigma_{IT}$ phase space (see **supplementary text** for details). We initiate each trajectory at the peak of a Gaussian inhomogeneous site energy distribution. Therefore, at long times the energy loss $\Delta E$ (**Supplementary Figure 10**) over the course of a trajectory generally restricts the exciton to a slice of inhomogeneous site energy (and spatial) distribution with a diminished density of states relative to the initial condition.[7,15,30] We classify the nature of migration trajectories according to the power $\alpha$ that embodies the temporal evolution of their mean square displacement, $MSD \propto t^\alpha$. Although the use of a power law functional form is phenomenological, we find that it fits the simulation data well and with fewer parameters than an exponential decaying asymptotically to a finite value. $\alpha$'s deviation below unity signifies subdiffusive (sublinear with respect to time) mean square displacement. An example of both a linear and a sublinear trajectory's mean square displacement illustrates the distinction between these behaviors (**Figure**

**3b,c**). Averaging over many trajectories for a given point in the phase space enables us to map contour lines for the extent ($L_d$) and nature ($\alpha$) of exciton migration. The resulting overlaid contour plots of $L_d$ (grey shades) and $\alpha$ (red shades) are shown in **Figure 3a**, and we analyze their behavior in more detail when considering the general implications of this framework below.

**Combining measurement and model.** Presently, to identify the nature of exciton migration in CN-PPV we search in the same phase space for the intersection of the $L_d$ result from our TRUSTED measurement (orange contour) and the average energetic relaxation over the exciton lifetime, Δ$E$, (spectral diffusion) obtained from the simulations, relative to the measured value of Δ[33] (purple contour, see also **Supplementary Figure 10**). The intersection of these two curves also crosses the constraint (yellow contour) imposed by the width of the absorption spectrum (**Figure 1b**), giving us further confidence in pinpointing CN-PPV at the location of the blue dot. Since this location in the phase space corresponds to $\alpha$=0.95, we conclude that exciton migration in CN-PPV is essentially diffusive and that the large migration length of 16 nm results from the combination of diffusive migration with the long, 5-ns, fluorescence lifetime. Microscopically, the intrinsic broadening of the site energy is sufficient to make many sites energetically indistinguishable in spite of substantial inhomogeneous broadening, but is not sufficient to annul the downhill energy transfer bias imposed by a relatively large intrinsic Stokes shift. This Stokes shift is therefore likely to be the limiting factor in the extent of migration in CN-PPV solids.

**Role of material energy scales in exciton migration.** In addition to the particular analysis for CN-PPV, the constraints of our experiment and theoretical framework enable more general prediction of how the nature and extent of exciton migration vary as a function of the energy relationships that we have identified. The $L_d$ contours in **Figure 3** roughly fall along the antidiagonal with the highest $L_d$ values in the bottom left corner, where sites are essentially energetically indistinguishable. The $\alpha$ contours also roughly fall along the antidiagonal, with diffusive behavior found in the bottom left and subdiffusive behavior in the top right of the plot. In the limit of low site distinguishability and low downhill bias (bottom left) there are a high density of available sites for an exciton to hop to, and rapid hopping between them is possible, with little energy loss. This combination allows for extended diffusive migration ($\alpha = 1$). At the opposite extremes of high site distinguishability and large downhill hopping bias, each hop reduces the energy of the exciton and the density of available sites for subsequent hops. As a result, each hop takes longer than the previous one, and migration slows over the exciton lifetime ($\alpha < 1$). The migration length is also limited under these conditions because the exciton reaches an equilibrium energy within a narrow range of the inhomogeneous distribution that has a relatively low density of accessible sites, resulting in slow migration or even trapping.[34] Interpolating between the bottom left and top right of **Figure 3a**, one can therefore rationalize that both $L_d$ and $\alpha$ decrease along the diagonal of the plot. The roughly antidiagonal nature of their contours implies that the importance of site distinguishability and downhill bias can trade off to produce a similar resulting extent or nature of migration. Interestingly, however, a single $L_d$ contour can intersect multiple values of $\alpha$. That one can obtain the same overall extent of migration via hopping trajectories with different rates of energy loss implies that changes in the spatioenergetic landscape can be compensated by the manner in which it is traversed (**Supplementary Figure 11**). As an overall prescription for long range exciton hopping trajectories, one should seek to minimize the Stokes shift and inhomogeneous spectral linewidths of an electronically-coupled material *relative to the intrinsic site broadening*. As a corollary, substantial intrinsic site broadening should be able to compensate for comparable site energy dispersity and reorganization energy.

Although the migration length values on **Figure 3** are specific to CN-PPV, these contours' trends should hold for other materials, modulated primarily in value by the fluorescence lifetime, chromophore density, oscillator strength, and any orientational anisotropies. We therefore hypothesize where other electronically-coupled systems are situated within the above framework. Other conjugated polymer solids likely fall into a similar part of the generalized phase space as CN-PPV because substantial intrinsic site broadening is able to compensate the other energies in the problem. These other semiconductors generally suffer from shorter migration lengths (~5-20 nm),[3,7–12] which we attribute primarily to their characteristically shorter fluorescence lifetimes, restricting the trajectory duration without compensating with faster hopping rates. In fact, given that the migration length depends not only on the fluorescence lifetime but also on this lifetime relative to the typical exciton hopping time, it is remarkable that CN-PPV's 25-fold increase in the lifetime over the canonical MEH-PPV only translates to a twofold increase in $L_d$.[35] By contrast to conjugated polymer solids, nanocrystal arrays generally generate subdiffusive exciton trajectories located further from the phase space origin, albeit with longer migration lengths (tens of nm).[15] We attribute this difference to the much longer nanocrystal excited state lifetimes. In spite of their characteristically very narrow intrinsic linewidths that generate subdiffusion from polydispersity in size and energy, the slower hopping rates that accompany larger site spacings must still enable a compensating number of hops within the lifetime. In photosynthetic light harvesting, individual pigment-protein complexes can induce substantial energy relaxation, yet hopping between 'identical' protein complexes is paradoxically rapid (~tens of ps).[17,18] Here, having a hierarchy of intra- versus inter-protein interchromophore length and energy scales may illustrate a mechanism for high efficiency exciton transport that is less well-captured by our hopping framework, even though inter-protein energy transfer is largely incoherent.

Although in the above analysis we localized conjugated polymer solids to a particular quadrant of the phase space in **Figure 3**, we wish to make more specific predictions of what TRUSTED could reveal about the nature of exciton migration in other prominent conjugated polymer solids. Interestingly, despite their lower bandgaps that result in higher nonradiative rates and shorter exciton lifetimes, push-pull polymer materials are able to achieve high internal quantum efficiencies.[36–38] Our TRUSTED approach could be used to determine the balance of exciton vs. charge transport that underlies their high efficiencies. Furthermore, our experiments and analysis could dissect how the pseudo-charge transfer nature of their excited states, their large backbone dipole moments, and their substantial structural disorder affect the electronic energy landscape and resulting exciton migration properties. We anticipate that migration lengths would be shorter than in CN-PPV due to the difference in fluorescence lifetime but that intrinsic broadening might similarly be able to counterbalance the effects of disorder. The migration lengths would likely also be shorter in canonical amorphous polymer films such as in regio-random P3HT on account of its shorter lifetime. We hypothesize that TRUSTED could also reveal the combination of structural order, delocalization, and energetic broadening contributions that allow regio-regular P3HT to display an exciton diffusion length substantially longer than that of regio-random P3HT[39] and much more similar to that of CN-PPV. The fact that both partially ordered regio-regular P3HT and amorphous CN-PPV films support similar diffusion lengths highlights the complexity and multiplicity of material spatioenergetic landscape parameters that determine the observed diffusion length. Excitingly, TRUSTED should allow us to describe the combinations of the energy scales and morphologies that give rise to the diffusion lengths observed in different materials, even if these diffusion lengths appear similar due to different balances of competing contributions.

In sum, we have devised and executed a new, all-optical scheme to measure exciton migration within sub-diffraction excitation volumes on its native nanometer and picosecond scales. Through a combination of our measurements and simulations we determined that the disordered CN-PPV films that we interrogated exhibit a considerable exciton migration extent of ~16 nm in the diffusive regime, largely thanks to a relatively long fluorescence lifetime and to the intrinsic broadening of the chromophore site energy. In addition to our measurement and analysis of exciton migration in CN-PPV films, we developed a more general framework in which to contextualize our results by distinguishing between the inter- and intramolecular energy scales that influence the character and extent of exciton migration. We emphasize the significance of assessing inhomogeneous broadening and intrinsic chromophore Stokes shift effects on migration *relative* to intrinsic variations in chromophore site energies. As such, intrinsic site energy fluctuations are partially able to—or could be designed to—compensate for the latter effects in disordered electronically-coupled molecular systems. For example, deliberately enhancing intrinsic chromophore energy fluctuations on molecular (or even material) scales by design could become an important strategy to extend exciton migration in photovoltaics, while suppressing it could prevent degradation in modern organic displays. We posit that the additional levels of multiscale hierarchy in photosynthetic light harvesting—namely intra- versus inter-protein exciton transfer and a potentially-active protein scaffold with complementary physical properties to those of the pigment chromophores—could explain their exemplary transport efficiencies, which should be amenable to TRUSTED investigation in the future. Although the measured CN-PPV films appear, not surprisingly, to be amorphous on the scale of our measurement (**Supplementary Figure 12**), TRUSTED is also inherently amenable to resolving spatial heterogeneity in exciton migration. Comparing migration heterogeneity maps to those of the physical heterogeneities observed in complex material microstructure should be a powerful approach to elucidate correlations between advantageous physical and functional intermolecular configurations in many electronically-coupled molecular materials.

**Methods**

*Sample preparation:* Thin films of CN-PPV were prepared by spin-casting a ~2.5 mg/mL solution of CN-PPV in chloroform on glass microscope coverslips in a nitrogen glove box, where they were encapsulated to protect them from oxygen during measurement. The solution was prepared by dissolving 9.8 mg CN-PPV in 1 mL chloroform and stirring on a hot plate overnight, then diluting 0.255 mL of this solution with 0.745 mL chloroform. The resulting solution was heated to 50° for ~4 hours before spin-casting at 8000 RPM for 1 min with an acceleration of 8000 RPM/sec.

*Characterization:* A representative absorbance spectrum of a CN-PPV film was acquired with a UV−vis spectrophotometer (Agilent Cary 100). The fluorescence spectrum of CN-PPV solids was obtained with a Horiba Fluoromax-4 fluorimeter with 540 nm excitation wavelength and a 5 nm detection slit width.

*Migration Measurements:* The TRUSTED scheme described above was employed in a home built confocal microscope with a 63× 1.4NA Plan Apo Leica objective (HC PL APO 63x/1.40 oil CS2, Leica Material #11506350). The excitation and depletion laser pulse trains at 200 kHz were derived from third-harmonic and second-harmonic noncollinear optical parameteric amplifiers (NOPA) (Light Conversion), respectively, pumped by a 10 W Light Conversion PHAROS regeneratively amplified laser system with a fundamental wavelength of 1030 nm. The 3 fJ excitation pulse was centered at 540 nm, and the two 240 pJ depletion pulses were centered at 740 nm with a bandwidth set to 14 nm. The depletion pulses were produced, with a variable relative delay, by splitting the initial pulse with a polarization beam splitting cube and directing one pulse though an optical delay stage (Newport ILS-250-CC) before recombining the pulses with a second polarization beam splitting cube. Both the pump and depletion pulses were fiber coupled into single mode polarization maintaining fibers to produce high quality Gaussian modes. A vortex phase mask (RPC Photonics VPP-1a) was then used to generate the annular depletion pulse beam mode. The pulses were then directed into the microscope with dichroic mirrors (Chroma T650spxr and T600lpxr-UF2) through a quarter waveplate positioned to circularly polarize the depletion pulses. During the experiment, the sample is rastered with a PI Nano scanning piezoelectric stage (P-545.3C7) in steps of 30 µm over a 60 µm × 60 µm area. Data from the resulting nine spatial locations can be averaged to improve the signal to noise ratio or analyzed separately. Epifluorescence is collected between 635 and 640 nm through dichroic mirrors and emission filters (two ET625/30m and one ET640/10m from Chroma) and is focused onto a fast-gated SPAD detector with a 200 ps rise time (Prof. Alberto Tosi, SPAD lab, Politecnico di Milano; PicoQuant) controlled by a Picosecond Delayer (MPD) that is triggered just after the arrival of the second depletion pulse to eliminate fluorescence occurring before the definition of the detection volume. We phase lock the detection data stream to the timing of an optical chopper (Newport 3501) placed in the excitation pulse line, so that we may separately determine the photon count rates during the "excitation on" and "excitation off" chopper phases for multiple cycles. The count rates obtained during these open and closed phases of the chopper are each corrected for the classic pile-up effect with a simple Poisson correction factor (see Supplementary Information) before we take the difference of the two to isolate the count rate that is attributed to the modulated excitation pulse only. The second depletion pulse is separately modulated with a shutter so that data collected with this pulse blocked can be used as a reference and control. The signal vs delay time obtained when this second depletion pulse is unblocked is divided by the signal vs delay time observed when it is blocked. The resulting data is then normalized to the extrapolated value of this ratio at zero delay time to calculate the fraction of remaining excitations in the detection volume as a function of the delay time (see Supplementary information for more details).

*Fitting methods*: The length scale of exciton migration was extracted from the experimental data using a custom fitting function. The fitting function is a simplified model of the experiment, which employs diffusive expansion of the exciton distribution between its creation and detection. It explicitly includes the action of the modes of all three light pulses in the experiment (see Supplementary Information for details). The fit parameter is the diffusivity, $D$, which is then used to calculate the migration length, $L_d = \sqrt{D\tau}$. The uncertainty in the migration length is then found through a combination of the uncertainty in the diffusivity from the fit, which accounts for the error in the data and the quality of the fit, and the uncertainty in the resolution of the microscope, which is an input parameter for the fit (see Supplementary Information for details).

*Monte Carlo Simulations:* Simulations of incoherent exciton hopping trajectories were performed with discrete hops on an 80 nm x 80 nm 2D hexagonal lattice with periodic boundary conditions and a density of 1.4 sites/nm$^2$. For each trajectory, site energies were randomly assigned in accordance with the assumed Gaussian inhomogeneous broadening width, and trajectories were initiated at the average energy (peak of the inhomogeneous distribution). Hopping rates between pairs of sites were then calculated from the site energies, the site-to-site separation, the intrinsic spectral width of the sites, and the intrinsic Stokes shift, in a combined Dexter and FRET transfer model (See Supplementary Information for details). Approximately 1000 trajectories were averaged for each set of parameters and the resulting mean squared displacement over time was fit to a power law model to extract the migration length and characteristic power, $\alpha$.

Figures and Figure legends:

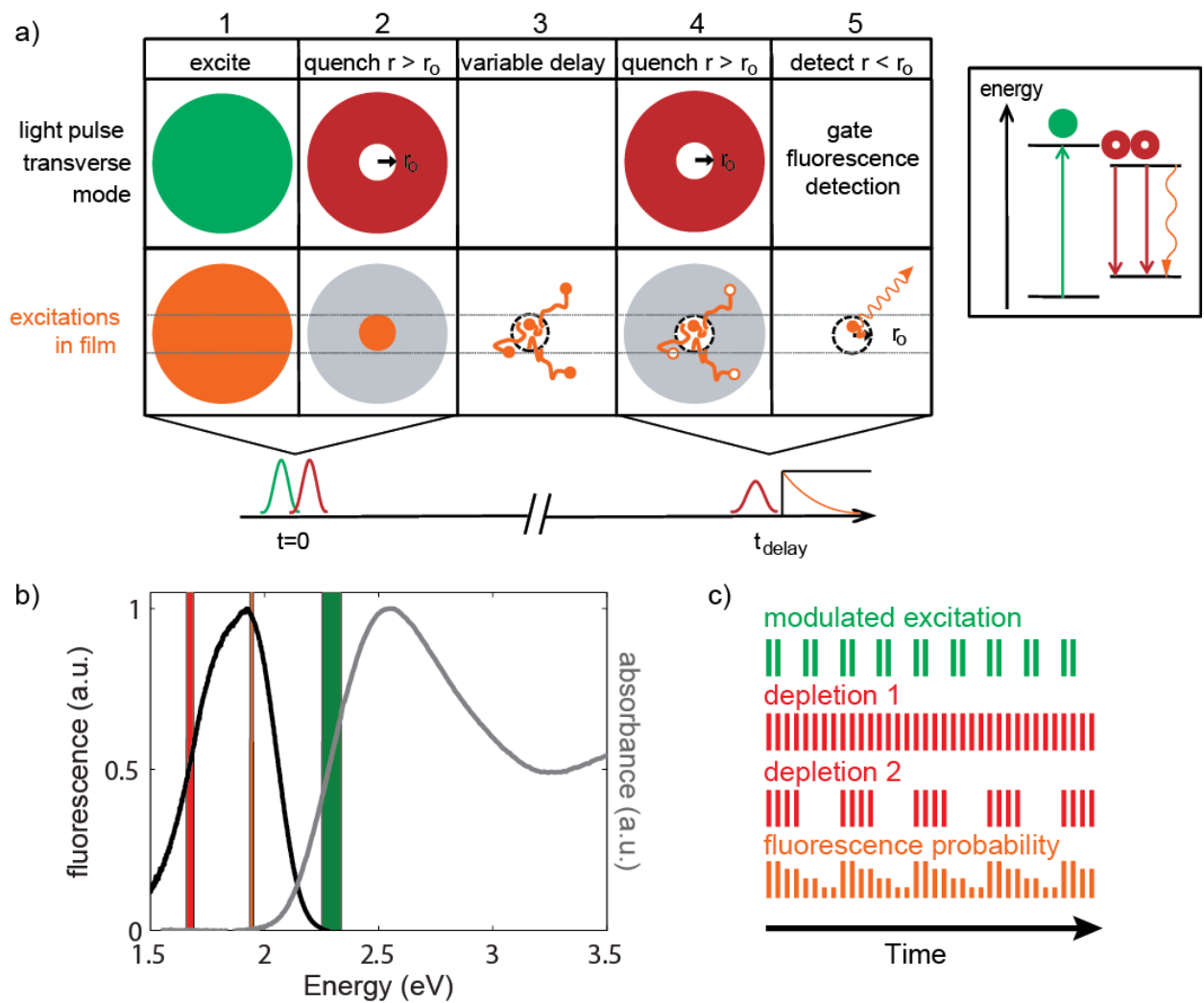

**Figure 1.** (a) TRUSTED sequence schematic showing the process in space, energy, and time. (b) Absorbance and fluorescence spectra of CN-PPV solids. Colored bands indicate the location and width of the pump pulse (green), depletion pulse (red), and fluorescence detection window (orange). (c) Sketch of modulation scheme employed. Only a few pulses are shown per on/off phase although the 200 kHz pulse repetition rate far exceeds the mechanical modulation frequencies.

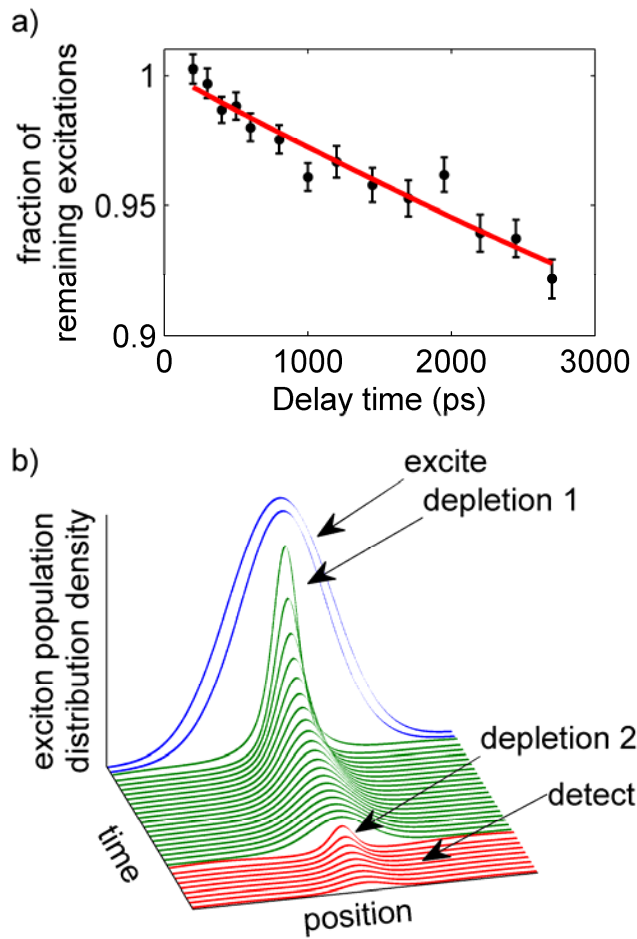

**Figure 2.** (a) Measured fraction of remaining excitation as a function of the delay time defining sub-diffraction excitation and detection volumes in the TRUSTED measurement. See **Figure S4** for raw data. Error bars are standard error of the mean. Red curve indicates a fit of the data to a exciton migration model to yield $L_d$=16 ± 2 nm. (b) Illustration of the components of the model, which includes the diffusive expansion of the initial exciton distribution, subject to the spatially varying depletion pulses.

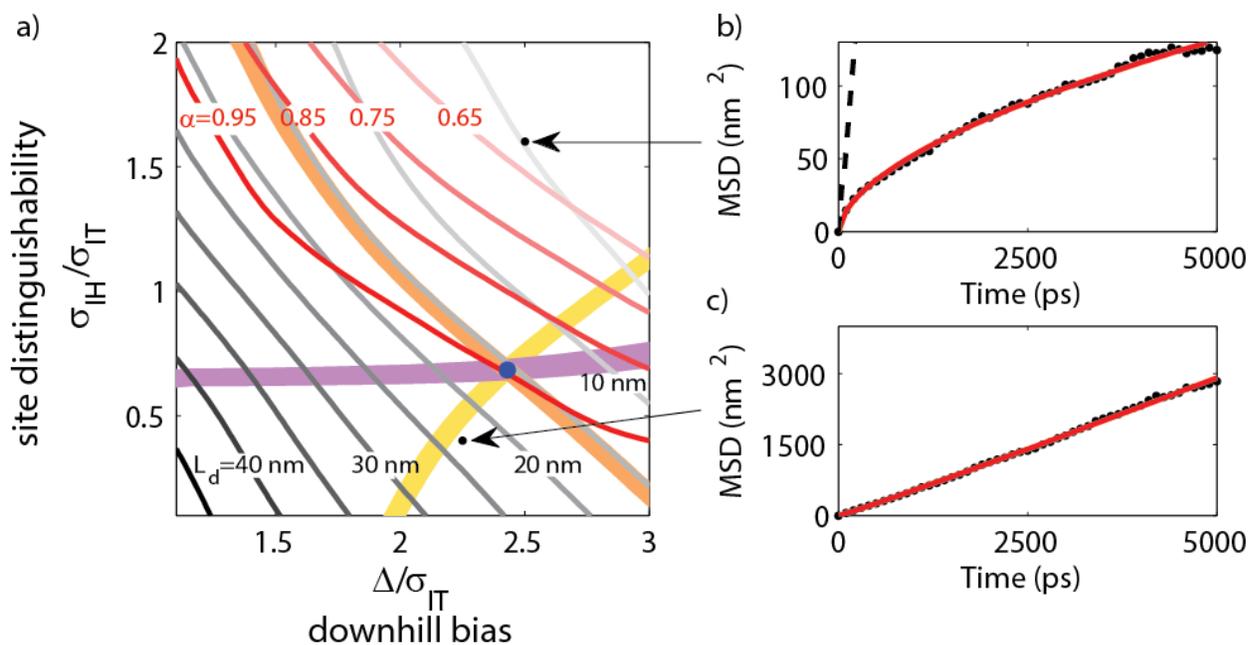

**Figure 3.** (a) Monte Carlo exciton hopping simulations results plotted in the $\sigma_{IH}/\sigma_{IT}$—$\Delta/\sigma_{IT}$ phase space, where $\sigma_{IH}/\sigma_{IT}$ represents the effective distinguishability of chromophore sites and where $\Delta/\sigma_{IT}$ represents the downhill hopping bias. Contours, obtained after averaging trajectories at each point in the phase space, for $L_d$ are shown in grey shades; contours for $\alpha$ are shown in red shades. The location of the CN-PPV solid in the phase space is indicated with the blue dot. It is found by obtaining the intersection of the TRUSTED $L_d$ result (orange contour), the contour for average trajectory energy loss relative to measured intrinsic Stokes shift (purple contour), and a contour obtained through the constraint on $\sigma_{IH}$ and $\sigma_{IT}$ imposed by the measured width of the absorption spectrum (yellow contour). The width of each of these three contours corresponds to a 1-$\sigma$ uncertainty. (b) and (c) Example mean square displacement vs time for the two points indicated with arrows on the phase space in (a). Data (black dots) is average of approximately 1000 trajectories, and fit to determine $\alpha$ is shown in red. Although (c) shows diffusive (linear) behavior for $\alpha=1$, (b) corresponds to $\alpha=0.62$. The dashed line represents a tangent to the slope at zero time to emphasize the extent of subdiffusivity. Note the stark difference in scales between (b) and (c).

**Acknowledgements:** This work was supported by a David and Lucile Packard Fellowship for Science and Engineering to N.S.G., by The Dow Chemical Company under contract #244699, and by STROBE, A National Science Foundation Science & Technology Center under Grant No. DMR 1548924. We thank A. Tosi and M. Buttafava of SPAD lab, Politecnico di Milano, for discussions and the generous trial of the fast-gated SPAD and N. Bertone and PicoQuant GmbH for providing a demo of the HydraHarp400 photon counting apparatus. We thank D. M. Neumark for the use of a grating stretcher. S.B.P. acknowledges a Department of Energy Graduate Research Fellowship (contract no. DE-AC05-060R23100) and N.S.G. acknowledges an Alfred P. Sloan Research Fellowship and the Camille and Henry Dreyfus Teacher-Scholar Program.

**Author contributions:** S.B.P., L.D.S.G., and N.S.G. designed the research. S.B.P. and

# Supplementary Information for Resolving Ultrafast Exciton Migration in Organic Solids at the Nanoscale

Samuel B. Penwell, Lucas D. S. Ginsberg, Rodrigo Noriega, and Naomi S. Ginsberg

## Experimental details and data collection

### Determining the size of the excitation volume

The size of the excitation volume is determined by measuring the resolution of STED images (one depletion pulse) of a CN-PPV nanoparticle, as described in previous work.[1] The resulting resolution curve is shown in Supplementary Figure 1. The best excitation spot size yielded a full width at half max (FWHM) of 67 nm, while a typical depletion pulse energy used in our work of 240 pJ produced an excitation spot size of 85 nm FWHM. This resolution represents the effective size of the excitation volume created by the pump and first depletion pulse.

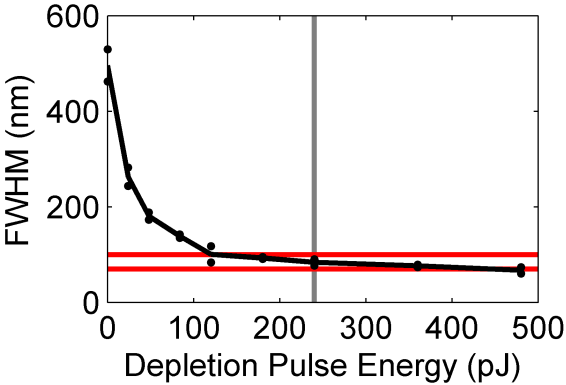

Supplementary Figure 1: The resolution (FWHM of the excitation volume) vs depletion pulse energy from imaging a CN-PPV nanoparticle. This figure is used in the fitting analysis to select the FWHM of the excitation volume at the depletion pulse energy used in the experiment (240 pJ as indicated in grey). The value of the FWHM at this depletion pulse energy is 85 nm. The red lines indicate FWHM values of 70 and 100 nm, which we take as a 95% confidence interval.

# Description of the experimental setup

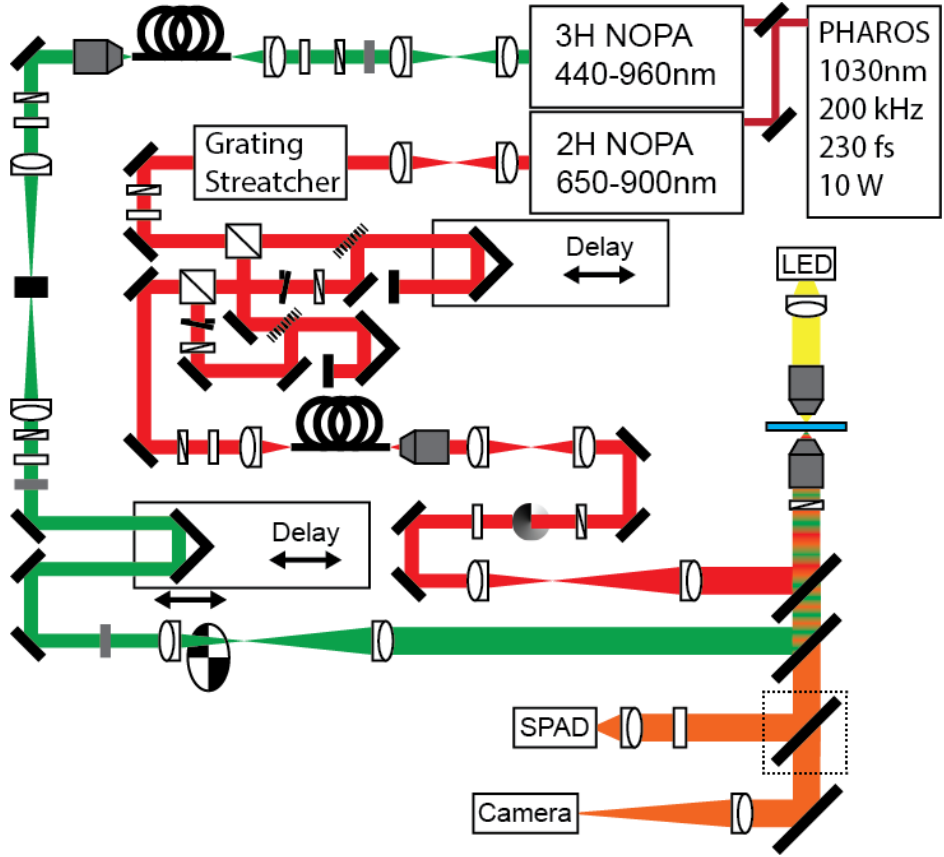

Supplementary Figure 2: Schematic of the experimental setup.

A schematic of the experimental setup is shown in Supplementary Figure 2. We use a PHAROS laser from Light Conversion, which is a ytterbium-doped potassium gadolinium tungstate (Yb:KGd(WO$_4$)$_2$) regeneratively amplified laser, which produces 230 fs pulses at 200 kHz with 50 $\mu$J/pulse and a 10 W average power. The PHAROS is used to pump two ORPHEPUS-N non-colinear optical parametric amplifiers (NOPAs), also from Light Conversion. The NOPA that produces the pump beam uses the third harmonic (3H) of the fundamental and the NOPA that produces our depletion pulse uses the second harmonic (2H) of the fundamental. For the experiments presented here, the 3H NOPA generated our pump pulse at 540 nm with a 20 nm bandwidth. Typical powers out of the NOPA where around 50 mW, or 250 nJ/pulse. The 2H NOPA generated our depletion pulse at 740 nm with a 14 nm bandwidth.

The pump pulse is collimated and then fiber coupled in to a single mode polarization maintaining fiber (ThorLabs PM105953) with a 10 mm focal length achromatic lens (ThorLabs AC080-010-A-ML). The Fiber output is collimated with a 10×/0.3 objective (Olympus UPlanFL N 10x/0.3NA, ThroLabs RMS10x-PF). The power of the beam is then controlled by focusing through a liquid crystal noise eater (ThorLabs LCC3111L), which attenuates the power and reduces power fluctuations. The power out of the noise eater is further reduced by a waveplate/polarizer pair and an absorptive neutral density filter. The transmitted beam is then optically chopped with a Newport chopper (Model 3501), then telescoped by a factor of 2 with 75 and 150 mm achromatic lenses. The beam is then coupled into the microscope with a dichroic mirror from Chroma (T600lpxr-UF2).

The depletion pulse is collimated out of the 2H-NOPA and then passed through a grating stretcher (Clark-MXR, Inc.). The beam is then split with a polarizing beam splitter cube (Newport 10FC16PB.5) into two lines. One of these lines is passed through a folded delay stage (Newport ILS-250-CC) with a retroreflector (PLX OW-25-1E), while the other is passed through matching stationary optics. Both lines pass through motorized waveplates (Newport PR50CC) and shutters before being recombined in a second polarization beam splitting cube (Newport 10FC16PB.5). The beam is then coupled into a single mode polarization maintaining fiber (ThorLabs PM105605) with a 10 mm achromatic lens (ThroLabs AC080-010-B-ML). The fiber output is collimated with a 10×/0.3NA objective (Olympus UPlanFL N 10x/0.3NA, ThroLabs RMS10x-PF). The beam is then passed through a vortex phase mask (RPC Photonic VPP-1a) and waveplate/polarizer pair to control the beam power. The beam is then expanded by a factor of 2.14 with 35 and 75 mm achromatic lenses. The beam is then coupled into the microscope with a dichroic mirror from Chroma (T650spxr). Finally, just before the microscope objective, the beams are sent through a $\frac{\lambda}{4}$ waveplate (Tower Optics 4505-0190 A-25.4-B-.250-N4) which makes the depletion pulse circularly polarized with handedness matched to the phase mask.

The microscope uses a 63×1.4NA Plan Apo Leica objective (HC PL APO 63x/1.40NA oil CS2, Leica Material #11506350). The objective is held in a custom mount on a Newport XYZ stage (Newport VP-25XL-XYZL). The sample is held in a PI Nano scanning piezo stage (P-545.3C7). Fluorescence is collected through the objective and transmitted through the dichroic routing mirrors for the pump and depletion pulses, then either imaged on a camera (ThorLab DCC1545M) or focused onto a single photon counting avalanche photodiode (SPAD) (Prof. Alberto Tosi, SPAD lab, Politecnico di Milano; PicoQuant)[2] through optional filters.

The motion of the Newport motorized stages is controlled with a Newport motion controller (XPS-Q8). The SPAD is gated with a Picosecond Delayer (MPD) and has a ∼ 200 ps rise time. The gate delay, on time, and subsequent hold off period can all be controlled through computer interfaces. Typical values for the gate duration and hold off time are 20 ns and 300 ns respectively. The counts from the SPAD are sent through a home built inverting amplification circuit then to a counting card on our data acquisition card (DAQ) (National Instruments PCIe-6321 with BNC-2090A breakout board). The DAQ is also sent a reference signal from the chopper at twice the modulation frequency, which it uses to trigger acquisitions of the SPAD counts and bin them according to the chopper phase. This enables the pile-up correction and background subtraction.[1]

## Method of power stabilization

The full data collection procedure can take up to 24-48 hours, and over this time there can be some fluctuations in the output power of the NOPAs, or some drift in the beam pointing, which impacts the fiber coupling efficiency and thus the power incident on the sample. These effects are monitored and corrected for. The noise eater in the pump line acts to control and stabilize the pump pulse power. In addition, both arms of the depletion pulse interferometer paths include motorized waveplates just before the polarizing beam splitting cube that recombine the two depletion pulses, which act to independently control their transmitted powers and facilitate corrections for power fluctuations. For the second depletion pulse, this process is more complicated. The beam is difficult to perfectly collimate over the range of the delay stage positions, so the beam size at the fiber changes with the delay setting unless some calibrated path-dependent compensation is performed. This impacts the coupling efficiency through the fiber, and thus the fiber output power varies with the delay setting. This effect is mitigated by setting a waveplate calibration curve that adjusts the waveplate angle for each delay setting, changing the beam power to compensate for the change in the coupling efficiency.

## Order of scan operations

The order of operations during data collection is also important to consider, as it impacts the duration of the experiment and the fidelity of the data. Our multi-variable data acquisition process can be conceptualized as a set of nested for-loops, defining and iterating over the values of each parameter. The inner-most level is the pump modulation. This modulation is driven by an optical chopper at 500 Hz and is the fastest variable. The next level is the position of the sample stage in XYZ, which is driven by piezos. We alternate between a set of spatial positions to eliminate heating effects that otherwise occur. We then modulate the first and second depletion pulse shutters. These shutters have response times of $\sim$ 20 ms and $\sim$ 200 ms respectively. This gives eight possible beam on/off combinations (Supplementary Table 1) for all spatial locations. Next, we change the second depletion pulse delay stage position, which is the slowest stage. For this reason we vary the delay linearly, instead of randomizing the delay points, which would provide better fidelity at the cost of increased duration of data collection. Finally, we iterate this entire process, which we refer to as a "scan", to average the data and gather statistics for error analysis. We also pause after every 3-10 scans to refocus the light on the sample and reset the waveplate angles to set the powers of each beam and compensate for any fluctuations in the waveplate calibration curve (caused, for example, by drift in the beam pointing with changes in the room temperature), both of which are automated procedures.

## Point spread functions

The point spread function of the pump and depletion pulses were determined by measuring the reflected intensity from an 80 nm Au bead, which is raster scanned in the focal plan. The resulting intensity maps are shown in Supplementary Figure 3.

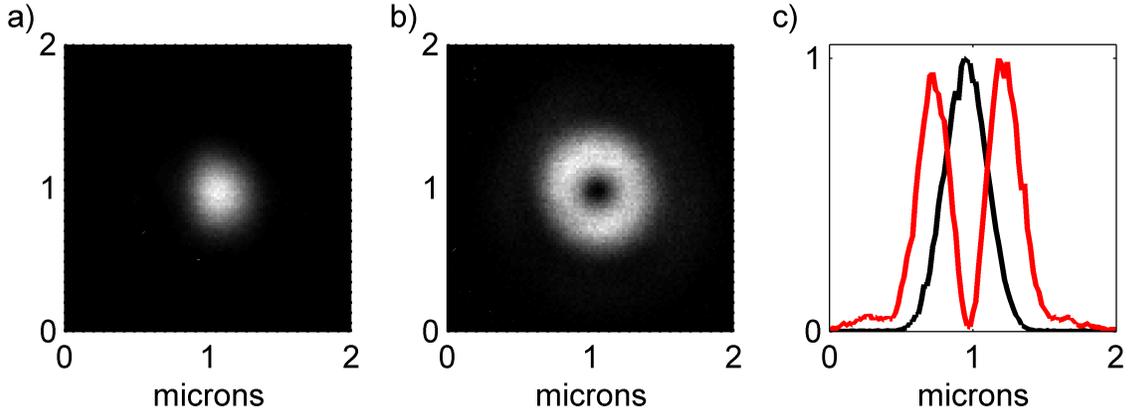

Supplementary Figure 3: The point spread function of a) the pump, b) the depletion pulses, and c) an overlay of normalized cross-sections through the pump and depletion point spread functions.

## Removal of 2PA background

The strong effect of 2PA in conjugated polymers presents a significant challenge to achieving super-resolution. The fluorescence that is induced by the STED pulse through 2PA needs to be removed so that the pump fluorescence can be isolated.[3] This process is similar to the case of transient absorption, where the small change in the probe transmission induced by the pump must be isolated from the large linear transmission of the sample. In the case of transient absorption, an optical chopper is used to modulate the pump pulse and a lock-in amplifier isolates the change in the probe intensity at the corresponding frequency. This approach fails here due deviations from linearity induced by the pile-up effect in single photon counting detectors, as outlined below. A correction can, however, be applied to account for the pile-up effect, restoring linear detector response and enabling the subtraction of the 2PA background through pump modulation.

Single photon detectors, like the single photon counting avalanche photodiode (SPAD) that we use to collect sample fluorescence, have a "dead time" after a detection event.[4] When a photon is detected by our SPAD it triggers an avalanche of charge that needs to be replenished before the detector regains sensitivity. This time scale is ∼50 ns for most detectors. We program our detector to be held in an off state for 300 ns after a detection event to avoid signal artifacts that can show up for short hold off times, such as after-pulses. Since typical fluorescent lifetimes are on the order of 1 ns, there will be at most one photon detected per excitation laser pulse. Thus it is not possible to detect a count rate higher than the repetition rate of the laser (200 kHz in our case). This effect introduces a non-linearity in the response of the detector to incident photons. This is a well known effect in single photon counting detectors and is referred to as the "pile-up" effect.[4, 5]

In order to correct for the pile-up effect, we need to understand the probability that multiple photons reach the detector from a given excitation pulse. Since there are many chromophores in excitation volume, this probability is governed by Poisson statistics, which say that if on average $\mu$ uncorrelated events are observed per time period, then the probability that $n$ events are observed in a given time period is described by the Poisson distribution,[4]

$$P(n,\mu) = \frac{\mu^n}{n!}e^{-\mu}. \qquad (1)$$

The single photon counting detector gives a binary response: either $n = 0$ or $n > 0$. We can therefore construct the probability that no photons are detected ($n = 0$) as one minus the ratio of the actually detected count rate to the repetition rate,

$$P(0,\mu) = 1 - \frac{\text{raw count rate}}{\text{repetition rate}} = e^{-\mu}. \qquad (2)$$

Then we can solve for the average,

$$\mu = -\ln\left(1 - \frac{\text{raw count rate}}{\text{repetition rate}}\right). \qquad (3)$$

This is the average number of photons incident on the detector from a single excitation pulse, so to convert back to a corrected count rate that describes the rate at which photons actually hit the detector even if they weren't detected, we multiply by the repetition rate.

$$\text{Corrected Count Rate} = -\ln\left(1 - \frac{\text{raw count rate}}{\text{repetition rate}}\right) \times (\text{repetition rate}). \qquad (4)$$

This simple correction can be used to dramatically extend the linear response of the detector. An example of an actual correction we performed is shown in Figure 4a.

Typically, isolating a signal from a background level can be achieved by modulating the signal and only looking at the component that changes with the correct frequency. This is often achieved with a lock-in amplifier. In this case, however, lock-in amplification fails to isolate the pump induced fluorescence due to the non-linearity induced by the pile-up effect on the detector. This effect can be overcome by incorporating the pile-up correction with the pump modulation scheme. The pile-up correction needs to be performed independently for both chopper phases (open and closed) in order to return both to a linear response regime before a subtraction can be performed to remove the background. This is achieved by sending the chopper phase reference to the computer so that photon detection events can be binned appropriately in software analysis. The pile-up correction can then be applied, according to Equation 4, to the measured count rates fro both the excitation pulse on and off cases. To yield the isolated count rate of pump-induced fluorescence we compute the difference between the two. This method works well to remove the STED-induced background due to 2PA (Figure 4b).

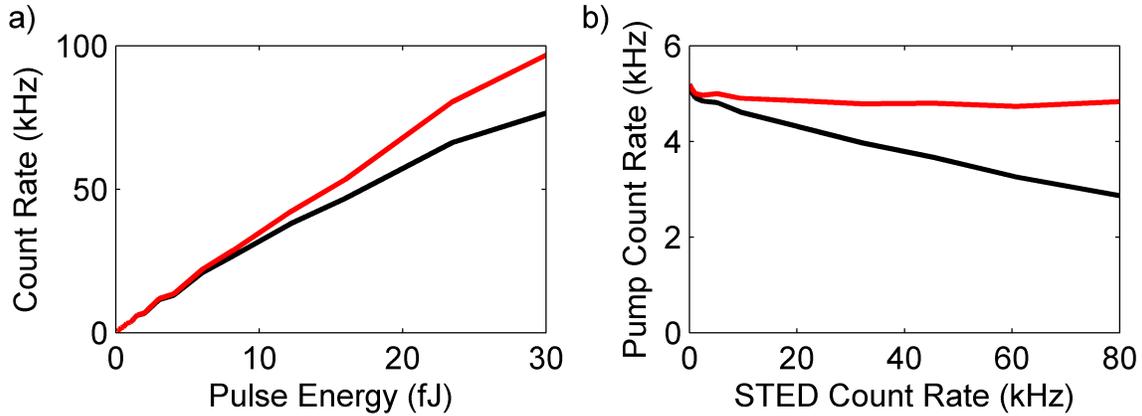

Supplementary Figure 4: Demonstration of the performance of the pile-up correction and modulation in removal of the 2PA background. a) Linearity of the detector response. Measured count rate, collected from the reflection of the pump pulse off a coverslip in the sample plane, vs the pulse energy with (red) and without (black) the pile-up correction. b) Isolation of pump-induced fluorescence from the 2PA background. The variation in the measured pump-induced count rate, for a constant pump pulse energy, as a function of the count rate of the STED-induced 2PA background with (red) and without (black) the pile-up correction.

## Modulation scheme

In this scheme the second depletion pulse is modulated at each gate position to calculate the normalized detection volume fluorescence. The pump pulse must also be modulated with an optical chopper to remove the background signal from fluorescence induced by the depletion pulse. In addition, the first depletion pulse is modulated to facilitate a control measurement. This results in 3 modulations with 8 possible beam combinations, outlined in Supplementary Table 1, all of which are collected for each delay position. Having access to all beam combinations allows for a large degree of freedom in the analysis, and plotting the various combinations can be informative.

| Name | Pump | depletion 1 | depletion 2 | Use | |
|---|---|---|---|---|---|
| All | 1 | 1 | 1 | Data | Migration |
| depletion 1 + depletion 2 | 0 | 1 | 1 | | |
| Pump + depletion 1 | 1 | 1 | 0 | Normalization | |
| depletion 1 only | 0 | 1 | 0 | | |
| Pump + depletion 2 | 1 | 0 | 1 | Data | Control |
| depletion 2 only | 0 | 0 | 1 | | |
| pump only | 1 | 0 | 0 | Normalization | |
| background | 0 | 0 | 0 | | |

Supplementary Table 1: Schematic of the modulation scheme for simultaneous acquisition of the normalization and control conditions.

# Data analysis

## Determining the fraction of excitations remaining

The quantity that underlies the sensitivity to migration is the variation in the fraction of the exciton population distribution quenched by the second depletion pulse for various delay times. The detected raw signal, however, is the fluorescence of the excitons. We determined that the best method to extract the fraction of the exciton population quenched by the second depletion pulse from the detected fluorescence involves two main factors. The first is to gate the detector on just after the action of the second depletion pulse for each delay time, removing sensitivity to any prior fluorescence. The second is to modulate the second depletion pulse, so that the fluorescence intensity with and without the application of the second depletion pulse, which generates the optical quenching boundary, may be recorded. The respective detected fluorescence levels are then proportional to the population of excitons in the sample with and without the action of the optical quenching boundary, so that the fraction quenched by the boundary can be extracted. This normalization eliminates the effects of fluorescence decay due to exciton relaxation. Furthermore, rather than considering the fraction of the exciton population that is quenched, we instead choose to consider the fraction that survives, as we feel it is better to plot the data in terms of the normalized detected fluorescence level. This ratio is referred to as the "normalized detection volume fluorescence".

$$\text{normalized detection volume fluorescence} = \frac{\text{detected fluorescence with depletion 2 on}}{\text{detected fluorescence with depletion 2 off}} \quad (5)$$

This quantity represents the fraction of the population that survives the action of the second depletion pulse at a given delay time, which is directly related to the spatial overlap of the second depletion pulse point spread function with the exciton population spatial distribution. Since the population distribution expands as migration proceeds, the spatial overlap with the second depletion pulse point spread function increases with the extent of migration. Thus, as migration proceeds, a smaller fraction of the population distribution will survive the action of the second depletion pulse, causing the normalized detection volume fluorescence to decrease with increasing delay time. Finally, the normalized detection volume fluorescence is normalized to its initial value to construct the "fraction of remaining excitations" in the detection volume.

## Raw signals and data processing

The experimental data was collected in the apparatus described above. A custom LabView interface was used to control all scan parameters and acquire the data. The pile-up correction and excitation modulation to remove the fluorescence induced by the depletion pulse is handled in LabView during data collection. The second harmonic of the 500 Hz chopping frequency is sent to the DAQ counting card to trigger the binning of counts from the detector. The count rate is computed for each bin and then sorted into two categories with every other bin corresponding to chopper open and closed phases. This yields two raw count rate data stream channels (Supplementary Figure 5c,d) one for each chopper state. These channels are averaged over multiple phase cycles to improve signal to noise and then corrected for the pileup effect[1] and subtracted to isolate the excitation pulse-induced fluorescence signal for the current pixel (Supplementary Figure 5a,b). This data is collected for all prescribed spatial locations and combinations of depletion pulse 1 on/off, depletion pulse 2 on/off, beam powers, and delay times and then this process is repeated for the prescribed number of scan iterations for averaging purposes. An example of these signals is shown in Supplementary Figure 5, where all spatial locations and scans have been averaged.

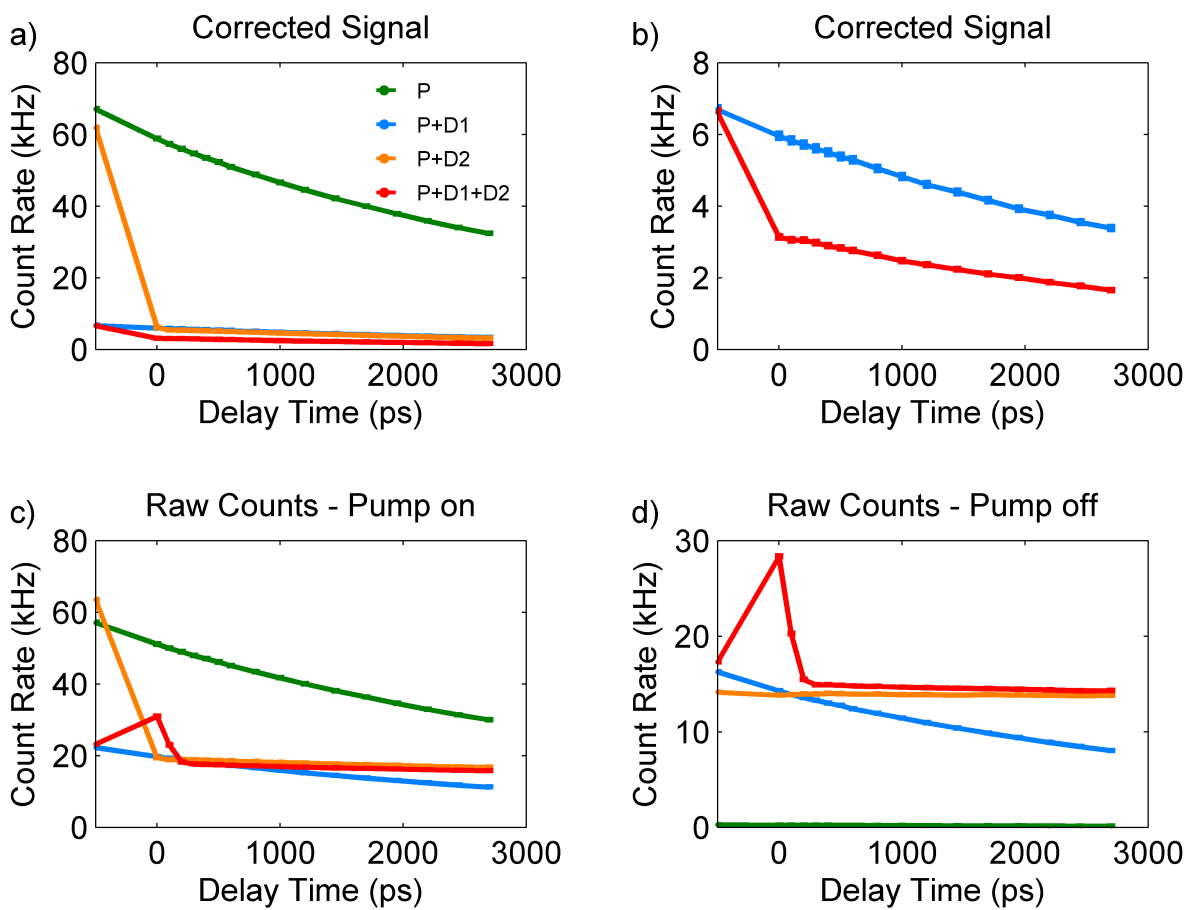

Supplementary Figure 5: Examples of the isolated pile-up corrected modulated signal and the underlying raw data. The graphs includes the overlaid data for the 4 possible depletion pulse combination, the excitation pulse (pump) only (green), pump + depletion 1 (cyan), pump + depletion 2 (orange), and pump + depletion 1 + depletion 2 (red). a) The isolated and pile-up corrected pump-induced fluorescence calculated from the raw data for the pump on and pump off chopper phases. b) same as (a) but only showing data for the cases that go into the determination of the fraction of remaining excitations. c) The raw count rates when the pump is on (chopper open). d) the raw count rates when the pump is off (chopper closed).

**Fitting methods**

Length scale for exciton migration were extracted from the experimental data using a custom fitting function. The fit function is based on a simplified model of the experiment and is illustrated in Supplementary Figure 6. The model assumes the initial population in the excited state is a Gaussian with a standard deviation obtained from fitting the point spread function of the excitation pulse (see Supplementary Figure 3a).

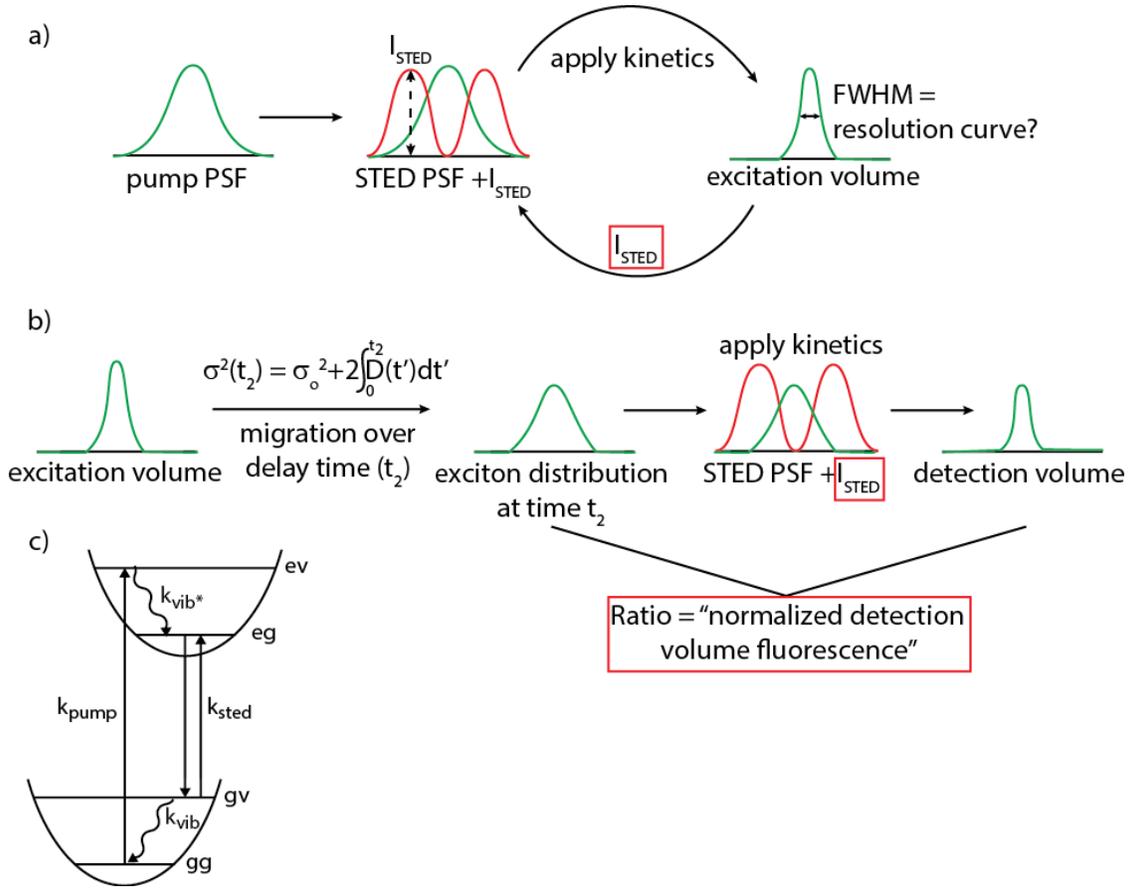

Supplementary Figure 6: A schematic of the the model used in the fitting routine. a) The first step starts with a Gaussian distribution of of excitons based on the excitation (pump) pulse point spread function, the kinetics of STED from Equation 6 are then applied where the radial dependence of the depletion field is taken from Equation 7 fit to the depletion pulse point spread function, and the intensity, $I_{STED}$ is variable. The appropriate value of $I_{STED}$ if found by fitting the FWHM of the resulting excitation volume to the observed value from the resolution curve in Supplementary Figure 1. b) The excitation volume found in the previous step is then propagated under migration over the delay time, $t_2$. The action of the second depletion pulse is then applied, with the same radial dependence and intensity as found for the first depletion pulse in the previous step. The normalized detection volume fluorescence is then the ratio for the excited state population after to before the action of the second depletion pulse. c) Energy level diagram, including ground (gg) and excited (gv) vibrational levels in the ground electronic state and ground (eg) and excited (ev) vibrational level in the excited electronic state, illustrating the kinetics of stimulated emission depletion, $k_{pump}$ is the rate constant for excitation, $k_{vib*}$ is the rate constant for vibration relaxation in the excited state, $k_{vib}$ is the rate constant for vibration relaxation in the ground state, and $k_{sted}$ is the rate constant for the transition driven by the STED field.

The kinetics of the first depletion pulse, illustrated in Figure 6c, are then treated analytically, under the assumptions of impulsive excitation ($k_{pump}$) and excited state vibrational relaxation ($k_{vib*}$), as described in Equation 6.

$$N_{eg}(t_{on}) = N_{eg}(0)\frac{e^{-\frac{1}{2}(k_{\text{eff}}+2k_{\text{sted}}+k_{\text{vib}})t_{on}}((1+e^{k_{\text{eff}}t_{on}})k_{eff} + (e^{k_{\text{eff}}t_{on}}-1)k_{\text{vib}})}{2k_{\text{eff}}} \quad (6)$$

Where $k_{\text{sted}}$ describes the rate at which transition driven by the depletion pulse approaches its equilibrium, $k_{\text{vib}}$ is the rate constant for the ground state vibrational relaxation that depletes the population of from the driven transition, $k_{\text{eff}} = \sqrt{4k_{\text{sted}}^2 + k_{\text{vib}}^2}$, $t_{on}$ is the duration of the depletion pulse, assumed to be a square wave, $N_{eg}(0)$ is the initial population in the excited state, and $N_{eg}(t_{on})$ is the population remaining in the excited state after the depletion pulse finishes. This expression is derived from solving the kinetics of stimulated emission deletion.

The depletion pulse point spread function radial dependence is assumed to have the functional form described by Equation 7[6],

$$I(r,\phi) \, \alpha \, r^2 \exp\left(-\frac{2r^2}{w_o^2}\right), \quad (7)$$

where $I$ is the intensity distribution, $r$ is the radius form the center of the beam mode, $\phi$ is the azimuth angle, and $w_o$ is the width, which is obtained from a fit to the experimental point spread function (see Supplementary Figure 3b). The ground state vibrational relaxation rate is assumed to be 0.05 ps$^{-1}$, the model is insensitive to this parameter as long as it is fast compared to the depletion pulse duration. The intensity of the first depletion field in the model is then set by performing a separate fit to find the intensity (Supplementary Figure 6a) that produces a confined distribution (excitation volume) of the desired FWHM, which is an input parameter set by the user. This value is chosen by measuring the resolution of an image of a CN-PPV nanoparticle with a depletion pulse of the same intensity used in the migration measurement. This resolution curve is shown in Supplementary Figure 1. This allows the model to account for the excitation and first depletion pulses and produce an initial distribution of the correct size. To proceed the model must account for the action of migration and the second depletion pulse (Supplementary Figure 6b). The excitation distribution from the previous step is assumed to be a Gaussian, which is a good assumption when the depletion pulse duration is long compared to the ground state vibrational relaxation. The migration of this Gaussian distribution is then propagated over the time delay with a diffusion model, where the the evolution of the Gaussian is described by $\sigma(t)^2 = \sigma_o^2 + 2Dt$. The second depletion pulse is assumed have an identical intensity and radial distribution as first depletion pulse and its action is again applied using Equation 6. The normalized detection volume fluorescence is then calculated as the ratio of the excited state population after to before the action of second depletion pulse, and is normalized to the initial value to calculate the fraction of remaining excitations.

This model represents a simplified simulation of the experiment which can then be fit to experimental data, using matlab's "lnsqcurvefit()" function. The parameters of the fit are the diffusivity, $D$, and an offset to correct the initial percentage quenching.

**Uncertainty in the fit**

This fitting procedure yields the parameters which define the time dependence of the diffusivity. The uncertainty on these parameters is then calculated based on the Jacobian of the fit, the experimental error, and the residuals following an established procedure.[7] The main goal is to transform the covariance matrix of the data ($C^d$) into a convariance matrix for the fit parameters ($C^p$), which can be expressed as:

$$C_{ij}^p = \frac{dp_i}{dy_k} C_{ks}^d \frac{dp_s}{dy_j} \quad (8)$$

or in matrix form as:

$$\mathbf{C^p} = \left[\frac{\mathbf{dp}}{\mathbf{dy}}\right] \mathbf{C^d} \left[\frac{\mathbf{dp}}{\mathbf{dy}}\right]^T \quad (9)$$

Where **p** is a vector of the parameters and **y** is a vector of the data points. It can be shown that:

$$\frac{d\mathbf{p}}{d\mathbf{y}} = (\mathbf{JJ}^T)^{-1}\mathbf{J} \qquad (10)$$

Where **J** is the Jacobian of the fit,

$$J_{ij} = \frac{df(x_j, \vec{p})}{dp_i}. \qquad (11)$$

Equation 9 can then be used to convert the covariance matrix of the data points to the covariance matrix of the fit parameters, where the data covariance is taken to be a diagonal matrix, with elements equal to the larger of either the square of the standard deviation or the square of the residual from the fit for each point. The resulting uncertainties on the parameters may then be propagated through the calculation of the diffusion length in the usual way.

This method accounts for the error in the data and the ability of the model to fit the data, but it assumes a known value for the degree of confinement, or resolution after the first depletion pulse. The uncertainty due to the choice of this value is illustrated in Supplementary Figure 7.

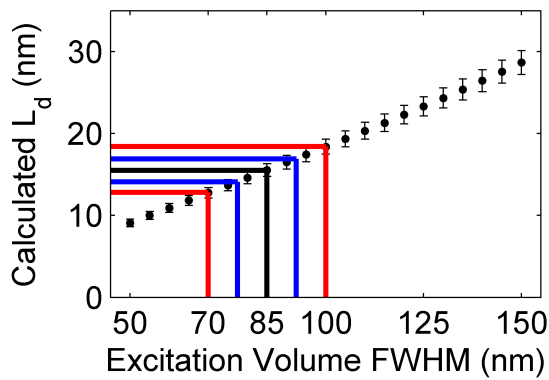

Supplementary Figure 7: The calculated $L_d$ that results results from running the fitting analysis on the experimental data for various values of the FWHM of the excitation volume. The error bars on the $L_d$ values account for the error in the quality of the fit. The black line indicates the result for our selected value of 85 nm. The red lines correspond to the 95% confidence range of 70 to 100 nm from Supplementary Figure 1, or $\pm 2\sigma$, and the blue lines indicate the corresponding results for $\pm 1\sigma$.

## The migration length

The diffusion length, $L_d$, is the length scale for the migration of particles with finite lifetimes. To derive its form, first consider the mean squared displacement ($MSD$) of an $n$-dimensional distribution centered at the origin with a diffusivity $D$:

$$MSD = 2nDt. \qquad (12)$$

The square of the diffusion length is then the time average of this quantity over the lifetime of the particles:

$$\begin{aligned} L_d^2 &= \langle 2nDt \rangle_t \\ L_d^2 &= \frac{\int_0^\infty 2nDt\, N(t)dt}{\int_0^\infty N(t)dt}, \end{aligned} \qquad (13)$$

where $N(t)$ describes the particle decay. In the special case where $N(t) = e^{-t/\tau}$, the integral has a simple solution:

$$L_d = \sqrt{2nD\tau}. \tag{14}$$

This quantity represents the average displacement of the particles from the origin over their lifetime, $\tau$. Another possible definition of $L_d$ is the average displacement in a particular direction $x_i$, which does not depend on dimensionality and is useful in describing the diffusion of particles towards a boundary along a given dimension:

$$L_d = \sqrt{2D\tau}. \tag{15}$$

A lot of the literature, related exciton diffusion in particular, drops both the dimensionality and the factor of two and defines the diffusion length as:[8, 9]

$$L_d = \sqrt{D\tau}. \tag{16}$$

Care must be taken to understand which definition is in use in a particular work. Although Equations 14 and 15 are more physically relevant, we will adhere to the commonly cited form in Equation 16, which implies that our reported values underestimate the typical micration distances by a factor of $\sqrt{2}$.

## Reproducibility of experimental results

Additional measurements of the exciton migration length in CN-PPV thin films are shown in Supplementary Figure 8a-e. The presented data were taken over a series of days, one set per day, at different positions on the sample. Note that each scan is an average over nine spatial locations spaced by 30 $\mu$m over a 60×60 $\mu$m$^2$ area, as in Supplementary Figure 2a. Also shown, in Supplementary Figure 8f, is a summary of the value of $L_d$ returned by the fit for each location, shown with the error from the fit, but not the error from the selection of the degree of confinement, which would likely be systematic in these successively collected results, for comparative purposes. The consistency in the reported value of $L_d$ over different days of data collection and different sample locations indicates that the measurement is reproducible.

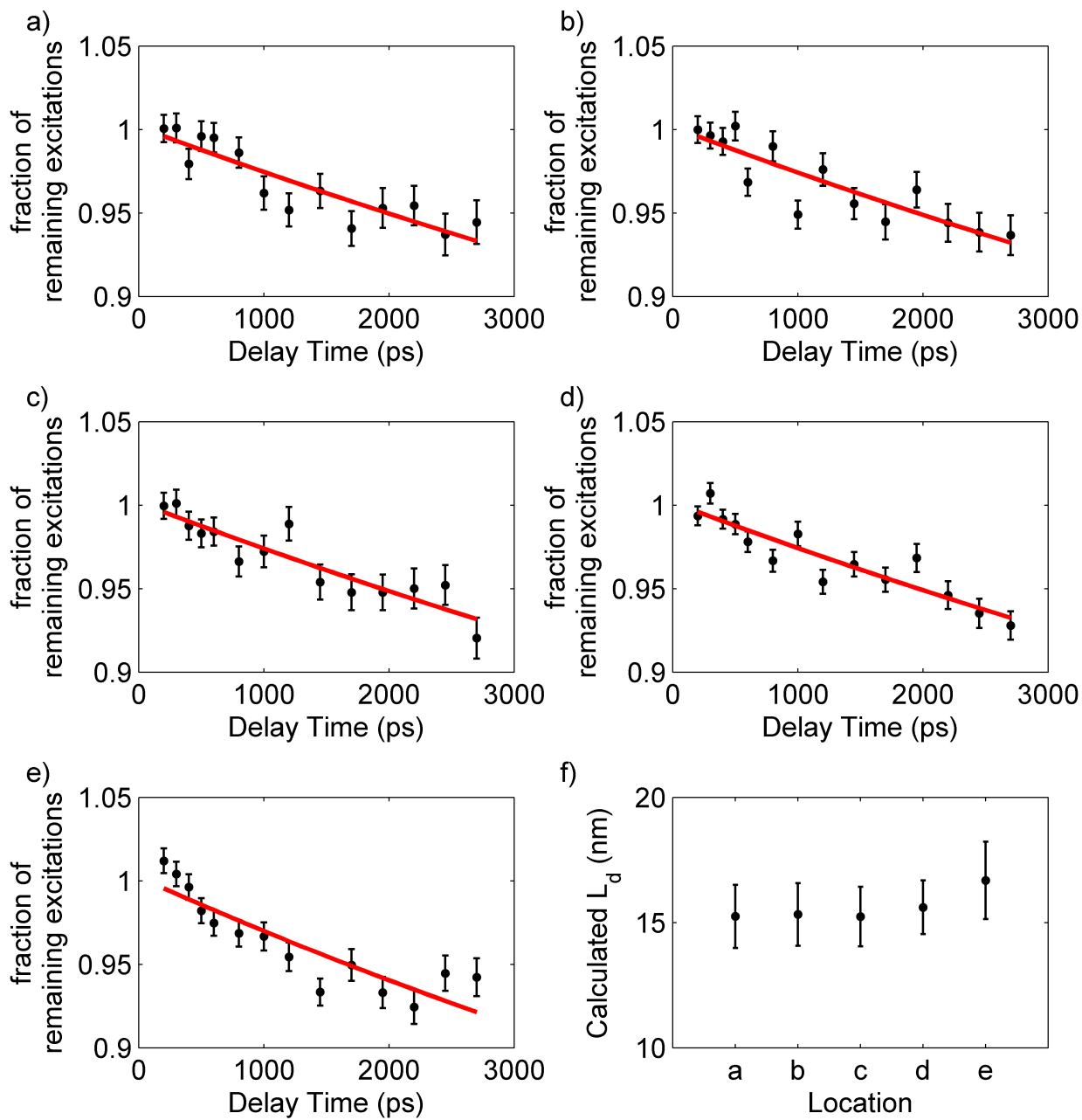

Supplementary Figure 8: Additional measurements of the exciton migration length in CN-PPV thin films, at sample coordinates relative to the center of the sample of: a) X = 2.72 mm Y = 2.80 mm, b) X = 3 mm Y = 2.88 mm, c) X = 2 mm Y = 1.88 mm, d) X = 1 mm Y = 0.88 mm, e) X=0 mm Y = 0.13 mm. f) A summary of the value of $L_d$ returned by the fit for each location.

## Excitation density dependence

Exciton annihilation could also produce a trend in the fraction of remaining excitations that would appear very similar to those produced by exciton migration. Annihilation, however, has an exciton density dependence, and migration does not. To determine the contribution of annihilation to the observed normalized detection volume fluorescence trend, we therefore performed a series of migration measurements at several excitation densities. The results are summarized in Supplementary Figure 9a, which shows no dependence of our calculated diffusion length over the range of excitation densities explored. Additionally, Supplementary Figure 9b shows the excitation density dependence of the fluorescence after the first depletion pulse (during migration), which is very linear over this range. In the absence of the first depletion pulse, which quenches a significant fraction ($\sim 90\%$) of the initially excited exciton population, the observed count rate is much higher and begins to saturate over this range of excitation energy, likely due to annihilation. The quenching effect of the first depletion pulse, however, reduces the exciton density to a linear regime and the saturation in the count rate is no longer observed. These results illustrate that the experiments are run in a linear regime where annihilation is negligible.

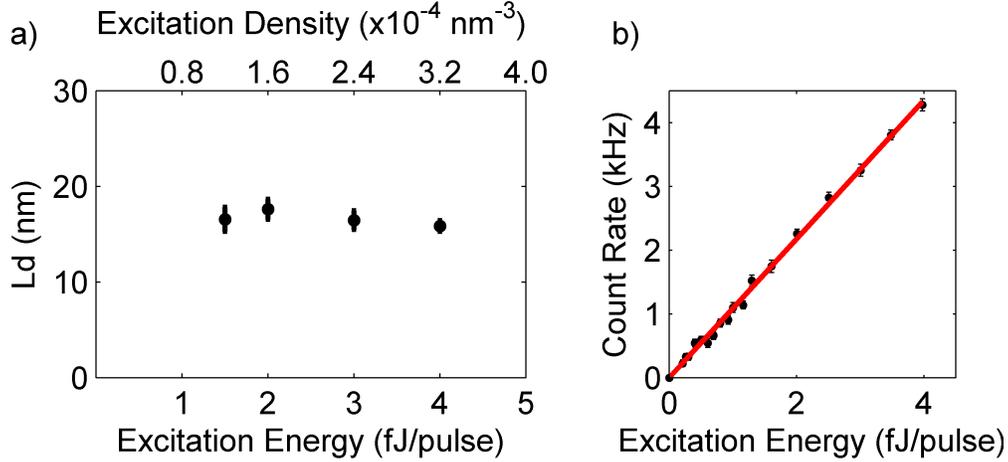

Supplementary Figure 9: Summary of the power dependence of our measurement. a) The excitation density dependence of our calculated diffusion length of CN-PPV. The bottom axis shows the excitation pulse energy used, while the top axis shows an estimation of the corresponding excitation density after the first depletion pulse. b) The variation of the observed count rate, when gating the detector on after the action of the excitation + depletion 1 pulses, vs the excitation energy.

## Monte Carlo simulations

### Description of simulation methods

To aid in the interpretation of our experimental results we have constructed a Monte Carlo simulation of incoherent exciton hopping.[10, 11] The goal of this simulation is to elucidate how the spatioenergetic and spectral parameters combine to determine the extent and character, diffusive or sub-diffusive, of migration. The simulation assumes discrete hops between sites on a $\sim$ 80 x 80 nm$^2$ 2D hexagonal lattice with periodic boundary conditions and a density of $\rho$=1.4 sites/nm$^2$. A 2D lattice is used for simplicity and computational efficiency, as there is no gradient in the $z$-axis in out experiments and the value of $L_d$, in the form commonly reported in literature, does not depend on dimensionality. For each random walk trajectory, the sites are randomly assigned absorptive transition energies from a Gaussian distribution, described by the inhomogeneous broadening, $\sigma_{IH}$, centered at the mean transition energy $\bar{\epsilon}$. The density of sites at a given energy is then,

$$\rho(\epsilon)d\epsilon = \frac{\rho}{\sqrt{2\pi}\sigma_{IH}} e^{-\frac{(\epsilon-\bar{\epsilon})^2}{2\sigma_{IH}^2}} d\epsilon. \tag{17}$$

The absorption and emission profile for each site are also assumed to be Gaussian, with widths determined by the site specific broadening, $\sigma_{IT}$, due to the Franck-Condon progression and fast thermal fluctuations. The normalized emission spectrum of a donor site of mean energy $\epsilon_d$ is,

$$D_{ems}(\epsilon, \epsilon_d) = \frac{1}{\sqrt{2\pi}\sigma_{IT}} e^{-\frac{(\epsilon-(\epsilon_d-\Delta))^2}{2\sigma_{IT}^2}} \tag{18}$$

where the Stokes shift, $\Delta$, accounts for the reorganization energy of an individual site (note that this quantity is not the same as the total observed Stokes shift from the film, which also includes the exciton energy relaxation due to migration to lower energy sites). While the absorption spectrum of an acceptor site of mean energy $\epsilon_a$ is

$$A_{abs}(\epsilon, \epsilon_a) = S(\epsilon_a) e^{-\frac{(\epsilon-\epsilon_a)^2}{2\sigma_{IT}^2}}, \tag{19}$$

where $S(\epsilon_a)$ is the oscillator strength at $\epsilon_a$. The rate of hopping between a donor and acceptor site is governed by Förster resonant energy transfer (FRET), for point dipoles given by[10, 12]

$$k_{FRET}(r, \epsilon_d, \epsilon_a) = \frac{1}{\tau} \left(\frac{R_o(\epsilon_d, \epsilon_a)}{r}\right)^6, \tag{20}$$

for

$$R_o(\epsilon_d, \epsilon_a) = \left[\frac{9c^4\hbar^4\eta\kappa^2}{8\pi n^4} \int \frac{A_{abs}(\epsilon, \epsilon_a) D_{ems}(\epsilon, \epsilon_d)}{\epsilon^4} d\epsilon\right]^{1/6}, \tag{21}$$

where $r$ is the distance between the sites, $\tau$ is the fluorescence lifetime, $R_o$ is the "FRET radius", $c$ is the speed of light, $\hbar$ is the reduced Planck's constant, $\eta$ is the quantum yield of fluorescence, $\kappa$ is a dipole orientation factor ($\kappa = 2/3$ for isotropic orientational averaging), $n$ is the index of refraction, and $\epsilon$ is energy.

This holds for long hops, however, short hops, over distances comparable to the chromophore size, are not described well by this equation, so our simulation switches from a the radial dependence of $r^{-6}$ to a Dexter-like exponential radial dependence for hops <2 nm, while constraining the overall radial dependence of the rate to be both continuous and smooth. To derive this radial dependence, we find the amplitude, $A$, and the radial decay constant, $c$ of an exponential function, $A e^{-cr}$ that matches the FRET rate and its derivative at $r = r_o$, where $r_o$ is set to 2 nm in our simulations. The constraints are,

$$A e^{-cr_o} = \frac{R_o(\epsilon_d, \epsilon_a)^6}{\tau} r_o^{-6} \tag{22}$$

$$\text{and} \quad -c A e^{-cr_o} = -6 \frac{R_o(\epsilon_d, \epsilon_a)^6}{\tau} r_o^{-7}, \tag{23}$$

which can be solved for $A$ and $c$, as

$$A = \frac{R_o(\epsilon_d, \epsilon_a)^6}{\tau} r_o^{-6} e^6 \tag{24}$$

$$\text{and} \quad c = 6/r_o. \tag{25}$$

By combining these expressions for $A$ and $c$ with the FRET rate, we can construct the overall radial dependence of the rates as,

$$k(r, \epsilon_d, \epsilon_a) = \begin{cases} \frac{1}{\tau}\left(\frac{R_o(\epsilon_d, \epsilon_a)}{r}\right)^6 & r > r_o \\ \frac{1}{\tau}\left(\frac{R_o(\epsilon_d, \epsilon_a)}{r_o}\right)^6 e^{-6\left(\frac{r}{r_o}-1\right)} & r \leq r_o \end{cases}. \tag{26}$$

For each trajectory that is run, the site energies are randomized and the excitation is placed on a site at the center of the density of states. The simulation then steps through a series of hops until the total elapsed time exceeds the lifetime of the trajectory, $\tau'$, which is selected for each trajectory from the probability distribution described by the observed lifetime, $\tau = 5000$ ps,

$$\tau' = -\tau \ln[\text{rand}()]. \tag{27}$$

The result of a given hop is calculated by first determining the transfer rates from the current donor site, to all other sites (acceptors) on the lattice,

$$\mathbf{K} = k(\mathbf{R}, \epsilon_d, \mathbf{E_a}) \tag{28}$$

where $\mathbf{R}$ is a vector of the distances to each site and $\mathbf{E_a}$ is a the vector of the corresponding the site energies. The site to hop to is selected by using the transfer rate vector, $\mathbf{K}$, as a probability distribution function

$$PDF = \frac{\mathbf{K}}{\text{sum}(\mathbf{K})}, \tag{29}$$

calculating the corresponding cumulative distribution function,

$$CDF = \text{cumsum}(PDF), \tag{30}$$

and selecting a value from this distribution as

$$ind = \text{find}(CDF >= \text{rand}(), 1, \text{'first'}), \tag{31}$$

where "find()" is the matlab function that returns one value that corresponds to the index $ind$ of the first site that is greater than or equal to the random value "rand()". The time for the hop to occur, $\Delta t$, is calculated from the total rate of transfer to all sites,

$$\Delta t = \frac{1}{\text{sum}(\mathbf{K})}. \tag{32}$$

The lattice is then re-centered around the selected location, using the periodic boundary conditions, effectively moving the lattice under the exciton instead of moving the exciton over the lattice.

The simulation tracks many parameters over the trajectory, such as the net displacement, the site energy, and the size and duration of each hop. Many trajectories (typically ~1000) are then combined to determine the average behavior of an exciton for a given set of site-specific and inhomogeneous broadening parameters and Stokes shift, producing graphs of the mean squared displacement (MSD or $\langle \Delta r^2 \rangle$) vs time and the average energy lost over the course of a trajectory ($\Delta E$). The migration length can be extracted from the average final MSD of all trajectories,

$$L_d^2 = \frac{\langle \Delta r^2 \rangle}{2n}. \tag{33}$$

The factor of $2n$, where $n$ is the dimensionality ($n = 2$ in the simulations here), has been divided out so that the reported $L_d$ value is consistent with the common practice in the literature of reporting migration lengths as $L_d = \sqrt{D\tau}$ rather than actual root mean squared displacement, RMSD $= \sqrt{2nD\tau}$.

Alternatively, the time dependence of the MSD can be fit to a functional form. For diffusive migration the MSD is linear in time,

$$\langle \Delta r^2 \rangle (t) = 2nDt, \tag{34}$$

where the slope is related to the diffusivity, $D$.

In real systems, however, the MSD often deviates from this trend. In these cases, the additional time dependence is assigned to the diffusivity,[11, 13, 14] such that

$$\langle \Delta r^2 \rangle (t) = 2n \int_0^t D(t')dt'. \tag{35}$$

Note that if $D(t)$ is constant, the diffusive equation is recovered. The most common functional form assumed for the diffusivity is a power law in time,

$$D(t) = D_o t^{\alpha-1} \tag{36}$$

so that the equation for the MSD becomes

$$\langle \Delta r^2 \rangle (t) = 2n \left(\frac{D_o}{\alpha}\right) t^{\alpha}, \tag{37}$$

where $\alpha = 1$ corresponds to the diffusive case, $\alpha < 1$ indicates subdiffusive behavior, and $\alpha > 1$ indicates superdiffusive behavior. The average MSD over the exciton lifetime is then,

$$\begin{aligned}\langle \Delta r^2 \rangle &= \frac{\int_0^\infty 2n \left(\frac{D_o}{\alpha}\right) t^{\alpha} e^{-t/\tau} dt}{\int_0^\infty e^{-t/\tau} dt} \\ &= 2n \left(\frac{D_o}{\alpha}\right) \tau^{\alpha} \end{aligned} \tag{38}$$

and the $L_d$ can then be calculated from Equation 33.

The result of the simulation is the final calculated migration length, and the nature of the migration, as characterized by the power $\alpha$. The simulation can then be repeated for multiple combinations of the $\sigma_{IH}$, $\sigma_{IT}$, and $\Delta$, to examine how these parameters determine the observed extent and character of migration.

### Classification of nature and extent of migration

To further understand the importance of $\sigma_{IH}/\sigma_{IT}$ and $\Delta/\sigma_{IT}$, as well as the microscopic origin of subdiffusive migration, consider the behavior of individual trajectories initiated at the center of the density of states at the level of individual hops. Broadly speaking there are four distinct cases, illustrated in Supplementary Figure 10, corresponding to the four quadrants in Figure 3a:

**Case 1 (Supplementary Figure 10b):** If both $\Delta/\sigma_{IT}$ and $\sigma_{IH}/\sigma_{IT}$ are large, there will be a large bias for hopping to acceptor sites lower in energy than the donor, and the sites will be distributed over a broad range of energies. In this case, the exciton will rapidly hop to spatially isolated states at the band edge and become trapped, due to the low density of accessible acceptor states at this energy and the very small isoenergetic spectral overlap that a large $\Delta$ value produces. Therefore, there is little to no migration ($L_d \sim 0$), there is a significant loss of energy ($\Delta E$), and the migration is subdiffusive $\alpha < 1$ since each step reduces the exciton energy, thereby reducing the density of available acceptors for the subsequent hop.

**Case 2 (Supplementary Figure 10d):** When $\Delta/\sigma_{IT}$ is large and $\sigma_{IH}/\sigma_{IT}$ is small, there is still a large bias for hopping to acceptor sites lower in energy than the donor, but now the entire collection of sites are confined to a narrow range of energies. The exciton will thus hop directly to the band edge (in ~1 hop), but the narrow distribution of acceptor site energies, relative the spectral width of each site, results in a relatively large density of available acceptors at this energy. The resulting migration will be diffusive ($\alpha = 1$) since all hops (after the first) are effectively equivalent, however, the diffusivity will be small due to the poor isoenergetic spectral overlap, so $L_d$ will be small but non-zero, and $\Delta E$ will be limited by the narrowness of the accessible band of states.

**Case 3 (Supplementary Figure 10a):** If $\Delta/\sigma_{IT}$ is small and $\sigma_{IH}/\sigma_{IT}$ is large, there will be only a slight bias for hopping to acceptor sites lower in energy than the donor and the sites will be distributed over a broad range of energies. In this case, the exciton loses a little energy with each hop, slowly approaching the band edge. Each hop also slightly reduces the density of accessible acceptor sites for subsequent hops, resulting in subdiffusive migration ($\alpha < 1$). An equilibrium is eventually reached where the large density of sites at higher energies balances the slight bias to move to lower energy sites or the exciton is trapped at a spatially and energetically isolated site. The overall energy lost in the trajectory will again be substantial, and the migration length will be moderate.

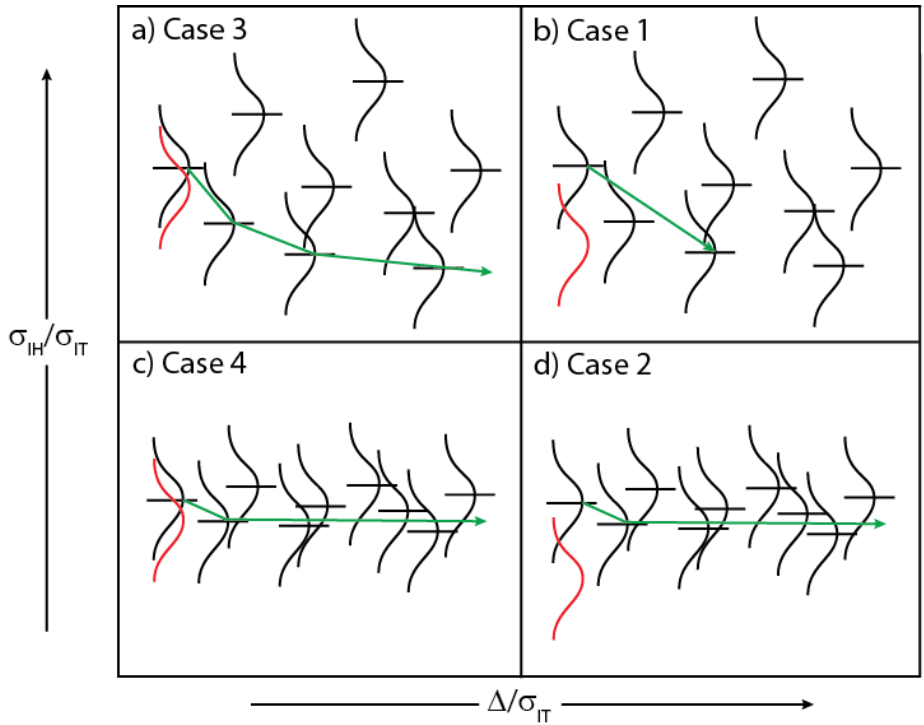

Supplementary Figure 10: Illustration of the microscopic trajectories for, a) case 3, b) case 1, c) case 4, and d) case 2 discussed in the supplementary text. The black curves indicate the absorption width of each site, the red curve is the emission profile of the first site to illustrate the size of the Stokes shift, the green arrow indicates the average progression of the trajectory. Note that while (c) and (d) appear similar, the time scales for migration will differ due to the difference in spectral overlap.

**Case 4 (Supplementary Figure 10c):** When both $\Delta/\sigma_{IT}$ and $\sigma_{IH}/\sigma_{IT}$ are small, there will be only a slight bias for hopping to acceptor sites lower in energy than the donor and the sites will be confined to a narrow range of energies. This narrow distribution of site energies, combined with the large degree of spectral overlap of isoenergetic sites (implied by the small value of $\Delta/\sigma_{IT}$), results in a high density of accessible acceptor states. The exciton will, thus, quickly ($\sim$1 hop) reach an equilibrium between the effects of the small bias to lower energies and the steep gradient in the density of states, and the corresponding energy loss with be small. Further hops will then be isoenergetic on average, resulting in diffusive migration ($\alpha = 1$), and the diffusivity will be high due to large spectral overlap of isoenergetic sites, resulting in a large migration length.

### Spectral diffusion or energy relaxation

In addition to the contours for $L_d$ and $\alpha$ that are plotted in Figure 3a of the main text, we are able to retrieve $\Delta E$, the average energetic relaxation in the hopping trajectories over the exciton lifetime. A map of -$\Delta E$ relative to the intrinsic Stokes shift $\Delta$ in the same phase space of $\sigma_{IH}/\sigma_{IT}$ and $\Delta/\sigma_{IT}$ is plotted as contours in red shades in Supplementary Figure 11 along with the contours for $L_d$ in grey shades. The average energetic relaxation obtained from the simulations that corresponds to the *film* Stokes shift of 0.65 eV obtained from the difference in peak energies of absorption and emission spectra in Figure 1b of the main text was used to localize the position of CN-PPV exciton migration on the plots in both Supplementary Figure 11 and Figure 3a. This blue spot in both figures is placed at the intersection of this average energetic loss contour and of the $L_d = 16$ nm contour, since this $L_d$ value is obtained from the TRUSTED measurements.

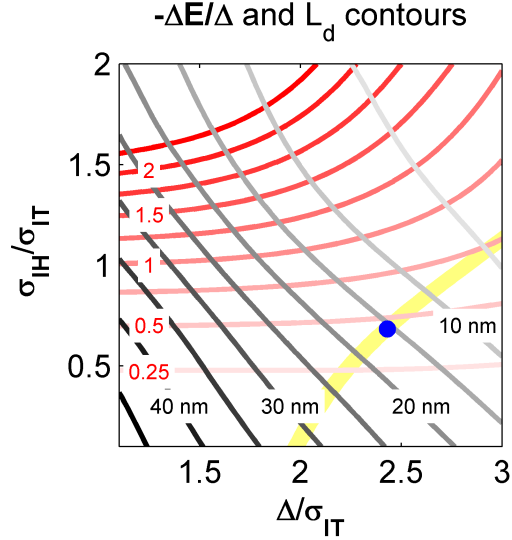

Supplementary Figure 11: An overlay of the contours of $L_d$ (grey) and $-\Delta E/\Delta$ (red) plotted against $\sigma_{IH}/\sigma_{IT}$ and $\Delta/\sigma_{IT}$. The yellow region represents the range of total broadening that is consistent with the observed absorption spectra, where the observed total broadening is the convolution of the inhomogeneous and intrinsic widths, $\sigma_{tot}^2 = \sigma_{IH}^2 + \sigma_{IT}^2$ for Gaussian distributions. The blue dot indicates the location of CN-PPV in the phase space.

## Multiple trajectories with same migration length

The same $L_d$ can be achieved for multiple values of $\alpha$, because the same net displacement over the lifetime can be realized for either an initially fast but subdiffusive or a slow but steady trajectory, as in Supplementary Figure 12. This variability in possible values of $\alpha$ for a given $L_d$ is pronounced for small $L_d$ but decreases as the $L_d$ increases, since there is a maximum allowed diffusivity (when $\sigma_{IT} >> (\sigma_{IH}+\Delta)$). Similarly, the same $\alpha$ can result in multiple $L_d$ values if the diffusivity is scaled, for instance by changing $\Delta/\sigma_{IT}$.

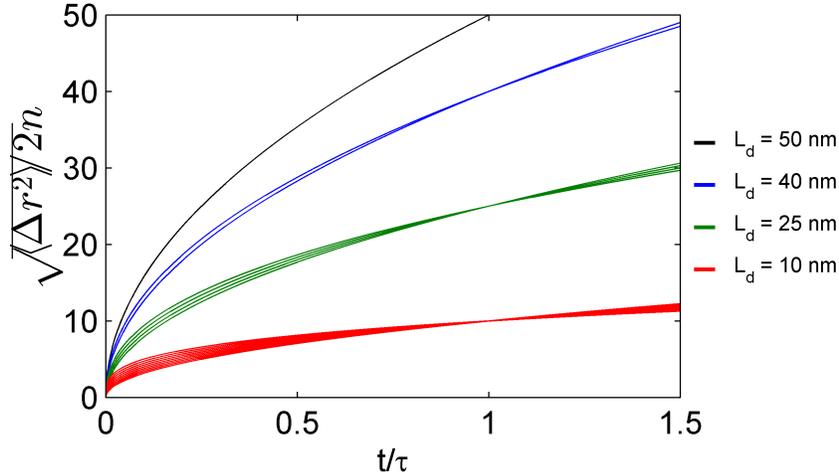

Supplementary Figure 12: Illustration of how multiple trajectories with different values of $\alpha$ can produce the same net displacement over the lifetime of the exciton and why this effect diminishes for large $L_d$ values. This plot assumes a maximum diffusivity that corresponds to $L_d = 50$ nm (black). Also plotted are some example trajectories that achieve $L_d = 40$ nm (blue), $L_d = 25$ nm (green), and $L_d = 10$ nm (red). As the target $L_d$ decreases, relative to the maximum value of 50 nm, a larger number of subdiffusive traces, with smaller values of $\alpha$, are possible.

## Without spatial averaging, migration in CN-PPV still appears uniform

The data for the nine individual sample locations, spaced by 30 $\mu$m over a 60×60 $\mu$m$^2$ area and collected as part of a single scan, which are averaged in the data reported in Figure 2a, are shown in Supplementary Figure 13a-i with overlaid fits in red. Note, that the error on each data point from individual sample locations is increased relative to Figure 2a due to the reduced averaging, however, the same general trend is seen. The variation in the calculated value of $L_d$ over these sample locations is shown in Supplementary Figure 13j with the error from the fit, but not the error due to the selection of the degree of confinement, which is a systematic effect in these simultaneously collected data sets. There does not appear to be any significant heterogeneity in the reported values of $L_d$ over the spatial locations, at least with the current degree of error in the data, implying that the sample is likely amorphous at the spatial resolution given by our single point measurements. It is possible, however, that further improvements to the experiment to increase the signal-to-noise ratio could reveal some underlying heterogeneity in this or other samples.

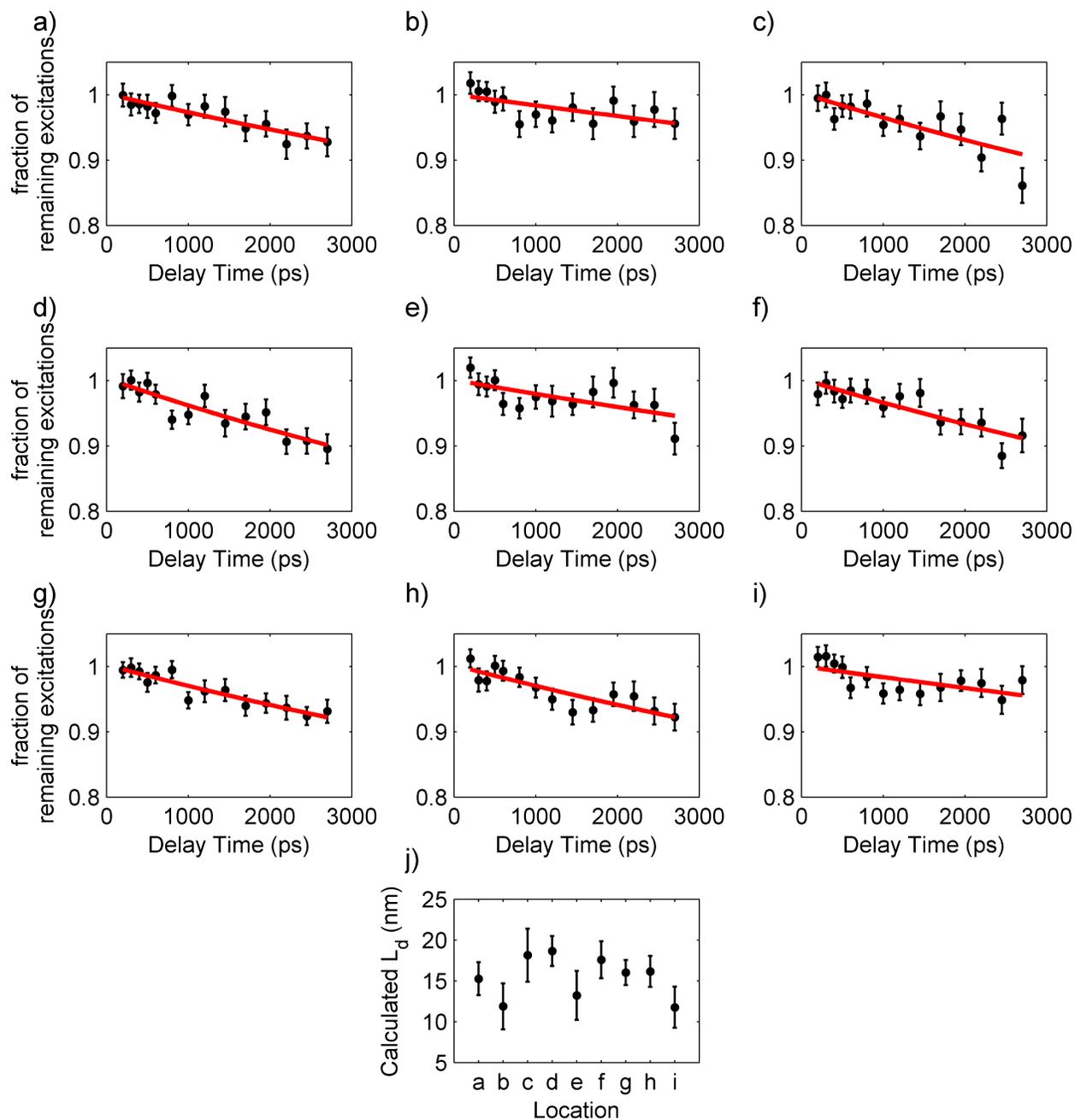

Supplementary Figure 13: The data from nine individual sample locations (parts a-i) collected as part of a single scan, which were averaged to produce the results shown in Figure 2a. Fits to the data are overlaid in red. Part (j) shows the variation in the calculated $L_d$ over these sample locations.

## Supplementary References